\documentclass[11pt,a4paper]{article}

\usepackage{url}
\usepackage{epsfig}
\usepackage[T1]{fontenc}
\usepackage[latin9]{inputenc}
\usepackage{multirow}
\usepackage{color}
\usepackage{placeins}
\usepackage{caption,amsmath,amssymb,amsfonts,graphicx,multirow, enumitem}
\usepackage{pstricks,pdflscape,morefloats,slashed}

\newcommand{\tb}{\bar t}

\newcommand{\tth}{ t \tb H}
\newcommand{\ttW}{ t \tb W}
\newcommand{\ttV}{ t \tb V}
\newcommand{\ttB}{ t \tb B}
\newcommand{\ttWp}{ t \tb W^{+}}
\newcommand{\ttWm}{ t \tb W^{-}}

\newcommand{\ttZ}{ t \tb Z}

\newcommand{\als}{\alpha_{\rm s}}
\newcommand{\shat}{\hat s}

\newcommand{\muf}{\mu_{\rm F}}
\newcommand{\mur}{\mu_{\rm R}}

\newcommand{\sigh}{\hat \sigma}

\newcommand{\nn}{\nonumber}
\newcommand{\tosv}{{\scriptscriptstyle \to}}

\RequirePackage[numbers,sort&compress]{natbib}

\def\beq{\begin{equation}}
\def\eeq{\end{equation}}
\def\bear{\begin{eqnarray}}
\def\eear{\end{eqnarray}}

\def\bet34{\beta_{kl}}

\begin{document}

\begin{flushright}
	MS-TP-20-01
\end{flushright}
\vspace{1cm}

\begin{center}
	{\Large\textbf{Associated top quark pair production with a heavy boson:
		\\[0.7em] differential cross sections at NLO+NNLL accuracy}}\\
	\vspace{1cm}
	Anna Kulesza$^{a,}$\footnote{\texttt{anna.kulesza@uni-muenster.de}}, Leszek Motyka$^{b,}$\footnote{\texttt{leszekm@th.if.uj.edu.pl}}, Daniel Schwartl\"ander$^{a,}$\footnote{\texttt{d\_schw20@uni-muenster.de}}, Tomasz Stebel$^{b,c,}$\footnote{\texttt{tomasz.stebel@uj.edu.pl}} and Vincent Theeuwes$^{d,}$\footnote{\texttt{vincent.theeuwes@uni-goettingen.de}}

	\vspace{.3cm}
	\textit{
		$^a$ Institute for Theoretical Physics, WWU M\"unster, D-48149 M\"unster, Germany\\
		$^b$ Institute of Physics, Jagiellonian University, S.\L{}ojasiewicza 11, 30-348 Krak\'ow, Poland\\
		$^c$ Institute of Nuclear Physics PAN, Radzikowskiego 152, 31-342 Krak\'ow, Poland\\
		$^d$ Institute for Theoretical Physics, Georg-August-University G\"ottingen, Friedrich-Hund-Platz 1, 37077 G\"ottingen, Germany
	}
\end{center}   

\vspace*{2cm}
\begin{abstract}
We present theoretical predictions for selected differential cross sections for the process $pp \to \ttB$ at the LHC, where $B$ can be a Higgs ($H$), a $Z$ or a $W$ boson. The predictions are calculated in the direct QCD framework up to the next-to-next-leading logarithmic (NNLL) accuracy and matched to the complete NLO results including QCD and electroweak effects. Additionally, results for the total cross sections are provided. The calculations deliver a significant improvement of the theoretical predictions, especially for the $\tth$ and the $\ttZ$ production. In these cases,  predictions for both the total and differential cross sections are remarkably stable with respect to the central scale choice and carry a substantially reduced scale uncertainty in comparison with the complete NLO predictions.

\end{abstract}

\clearpage
\tableofcontents
\setcounter{footnote}{0}

\section{Introduction}
\label{s:intro}

In the recent years the first two stages of the Large Hadron Collider (LHC) physics program, Run 1 and Run 2, have been completed with spectacular discovery of the Higgs boson, but with no indications of New Physics signals. In the future High Luminosity phase of the LHC, the luminosity will be increased by an order of magnitude.
Much larger statistical samples and better understanding of the systematic uncertainties will lead to a substantial reduction of experimental errors in the coming Run 3 and, specifically, at the High Luminosity LHC. In particular, one expects that for many important processes the experimental errors will get significantly smaller than estimated theoretical uncertainties. This context makes it absolutely necessary to maximize the theoretical precision of the Standard Model predictions for the processes measurable at the LHC. Together with increasing experimental accuracy this results in better knowledge of the Standard Model parameters, precise determination of the Higgs boson properties, and possibly to finding deviations from the Standard Model caused by new particles or interactions.

Studies of the most massive particles of the Standard Model: the top quark, the Higgs boson and the heavy electroweak gauge bosons: $W^{\pm}$ and $Z$ are of particular interest.  The Higgs boson sector is expected to exhibit enhanced sensitivity to New Physics effects since the measured value of the Higgs boson mass is highly unnatural within the Standard Model. Complementary, the Higgs boson and top sector are also crucial for solving the vital problem of the electroweak vacuum stability. The LHC energy and luminosity have allowed to search for and measure, for the first time, the processes of the associated production of the top-antitop quark pair with a heavy boson: the Higgs boson $H$ \cite{Aad:2014lma,Aad:2015gra,Khachatryan:2014qaa,Khachatryan:2015ila,Aad:2015iha,ATLAS:2016ldo,CMS:2016rnk}, 
$W^{\pm}$ or $Z$ \cite{Chatrchyan:2013qca,Khachatryan:2014ewa,Aad:2015eua,Khachatryan:2015sha,Aaboud:2016xve,Sirunyan:2017uzs,ATLAS:2018ekx,CMS:2019too}. These processes provide independent constraints of the electroweak couplings of the top quark and of the top quark Yukawa coupling. The most recent experimental analyses of $pp \to t\bar t W^{\pm}$ and $pp \to t \bar t Z$ were performed at $\sqrt{S} = 13$~TeV by the ATLAS \cite{ATLAS:2018ekx} and CMS \cite{CMS:2019too} collaborations,  using a data sample corresponding to a fraction of the collected Run~2 luminosity. In particular, the $Z$ boson's $p_T$ distribution was measured for the first time very recently~\cite{CMS:2019too}.

The current experimental accuracy~\cite{ATLAS:2018ekx, CMS:2019too} is already similar to the estimated theoretical uncertainty of the next-to-leading (NLO) QCD predictions for the  $t \bar t Z$ production and will only get better in the future measurements. Over the years a great deal of effort was invested to improve the theoretical description of the $pp \to \tth$ and $pp \to \ttV$ ($V=W^{\pm}, Z$) processes. The calculation of the NLO QCD corrections  ~\cite{Beenakker:2001rj, Reina:2001sf, Lazopoulos:2008de, Lazopoulos:2007bv,  Hirschi:2011pa, Kardos:2011na, Maltoni:2014zpa, Badger:2010mg, Campbell:2012dh, Rontsch:2014cca, Rontsch:2015una} was supplemented by their matching to partons showers~\cite{Frederix:2011zi, Garzelli:2011vp, Garzelli:2012bn, Alwall:2014hca, Maltoni:2015ena, Hartanto:2015uka}. The electroweak corrections and the combined electroweak corrections with the QCD corrections are also known~\cite{Frixione:2014qaa, Yu:2014cka, Frixione:2015zaa, Frederix:2017wme}. Furthermore, NLO QCD corrections were studied for production of the Z boson with off-shell top quarks~\cite{Bevilacqua:2019cvp}, while for the Higgs boson production with off-shell top quarks NLO QCD and NLO EW were calculated~\cite{Denner:2015yca, Denner:2016wet}. The dominant theoretical uncertainty comes from the higher orders of the QCD perturbative expansion and is estimated by variation of the factorization and renormalization scales. The fixed order results may be systematically improved by applying the soft gluon resummation technique. It is a rigorous approximation scheme based on the hard factorisation theorem within the leading twist collinear approximation. The soft gluon resummation picks up the leading contributions at all orders of QCD perturbative series enhanced by powers of logarithms with arguments depending on energies of soft gluons in higher order diagrams. The resummation is a standard tool to improve theoretical precision of fixed order calculations in QCD. The resummation is typically performed in one of the two frameworks: either using the direct QCD approach \cite{sterman, catani} or within the Soft Collinear Effective Theory (SCET) formulation \cite{SCET,Bauer:2001yt, Becher:2006nr, Beneke:2002ph}. 

The practical application of the soft gluon resummation to five-leg ($2\to 3$) particle processes with four legs carrying colour and with multiple colour channels has been developed in recent years within the direct QCD approach
\cite{Kulesza:2015vda,Kulesza:2016vnq,Kulesza:2017ukk, Kulesza:2017jqv,Kulesza:2017hoc, Kulesza:2018tqz, Kulesza:2019lye, Kulesza:2019adl} and using SCET \cite{Broggio:2015lya,Broggio:2016lfj,Broggio:2016zgg,Broggio:2017kzi, Broggio:2019ewu, Li:2014ula}. 
In the case of the associated $t\bar tH$ hadroproduction, first the direct QCD framework has been applied 
the absolute threshold limit at the improved next-to-leading logarithmic (NLL) accuracy \cite{Kulesza:2015vda}. In the following papers various approaches were developed \cite{Broggio:2015lya,Kulesza:2016vnq} to reach the full  next-to-next-to-leading logarithmic (NNLL) accuracy for the $t\bar t H$ process for the resummation performed in the invariant mass threshold limit \cite{Broggio:2016lfj,Kulesza:2017ukk}. 
A closely related analysis of soft gluon effects for a simpler $2 \to 3$ process, the $t\bar tW^{\pm}$ hadroproduction, 
started with resummation at the NNLL accuracy \cite{Li:2014ula}. Results at the NNLL accuracy 
were also obtained by other groups and calculations were extended for the associated $t \bar t Z$ production \cite{Kulesza:2017hoc, Kulesza:2018tqz, Kulesza:2019adl, Broggio:2016zgg, Broggio:2017kzi, Broggio:2019ewu}. 
Thus the NNLL resummation framework for $t\bar t B$ hadroproduction is available for all heavy bosons $B=H,Z$ or $W^{\pm}$.
Recently, it has been also proposed to improve the NLL treatment of the $t \bar t H$ cross section by resummation of the QCD Coulomb corrections in the absolute threshold resummation scheme \cite{Ju:2019lwp}.

The direct QCD and SCET frameworks are formally equivalent at a given logarithmic accuracy, but subleading terms beyond the formal accuracy may differ, depending for instance on the scale setting procedures. It is therefore important to compare the results obtained within both approaches in order to provide the highest level of theoretical reliability and to better understand theoretical uncertainties. So far the total cross sections for $t\bar t B$ hadroproduction at the LHC and the $t\bar t B$ invariant mass distributions were obtained through the NLO QCD~+~NNLL accuracy in both frameworks. In addition, using the SCET procedure a set of other differential distributions was calculated \cite{Broggio:2015lya,Broggio:2016lfj,Broggio:2016zgg,Broggio:2017kzi}. In the direct QCD approach, besides the invariant mass distribution of $t\bar t B$ \cite{Kulesza:2017ukk,Kulesza:2018tqz} only the $p_T$ distribution of the $Z$ boson was presented \cite{Kulesza:2019adl} to date. In this paper we fill this gap by providing the NLO~+~NNLL predictions of various kinematic distributions that are or may be measured in the $t\bar t B$ hadroproduction at the LHC.  Moreover we calculate several new distributions, that have not been considered in the literature yet. These are rapidity and azimuthal angle difference distributions of different pairs of the final state particles.
In all calculations presented here we include also electroweak effects up to the NLO accuracy. Calculations at NLO (QCD+EW) + NNLL accuracy in the direct QCD framework were first performed for the $p_T$ distribution of the $Z$ boson in $t \bar t Z$ hadroproduction \cite{Kulesza:2019adl}, and then for a set of differential $t\bar t B$ distributions in the SCET approach \cite{Broggio:2019ewu}. 
Also in the context of including electroweak effects this paper extends the results of \cite{Kulesza:2019adl} using the same approach in the direct QCD framework. Thus, the main goal of this analysis is to fully use the developed theoretical framework to provide the most accurate theory predictions for differential cross sections relevant for experimental measurements.  Special attention is given to the dependencies on various choices of central scales, and we show that the soft gluon resummation leads to high stability against the scale variations for all studied distributions. 

The paper is organized as follows: in Sec.\ 2 a short overview of the applied theoretical framework is given, in Sec.\ 3 the total and
differential cross sections through NLO (QCD + EW) + NNLL accuracy are presented for $pp \to t\bar t H$, $pp \to t\bar t W^{\pm}$ and  $pp \to t\bar t Z$,  and conclusions are given in Sec.\ 4.

\section{NNLL resummation}
\label{s:theory}

In the following, we use the formalism of threshold resummation in the invariant mass threshold limit, $Q^2 / \shat \to 1$, where $Q^2 =(p_t+p_{\bar t}+p_B)^2$ and $\hat s$ is the invariant mass squared of the colliding parton pair. Since the formalism has been developed and applied to the $pp \to \ttB$ process by us in the past~\cite{Kulesza:2016vnq,Kulesza:2017ukk, Kulesza:2017jqv,Kulesza:2017hoc,Kulesza:2018tqz, Kulesza:2019lye}, here we only present  the relevant key equations and refer the reader to our earlier work for a detailed description.

Resummation techniques allow to systematically take into account logarithmic terms  of the type $\als^n \left[ \log^r (1-z)/(1-z)\right]_+$, with $r \leq 2n-1$ and $z=Q^2/\shat$,  up to all orders in $\als$.  The resummation is performed in Mellin space, where Mellin moments are taken with respect to the hadronic threshold $\rho = Q^2/S$. The Mellin transform turns   the logarithms with arguments expressed in terms of the $z$ variable into logarithms of the Mellin moment $N$. The systematic inclusion of the logarithmic terms is performed through means of the factorized expression for the partonic differential cross section,
\begin{eqnarray}
\label{eq:res:fact}
\frac{d\tilde \sigh^{{\rm (NNLL)}}_{ij\tosv kl B}}{dQ^2}&&\hspace{-0.9cm}(N, Q^2,\{m^2\}, \muf^2,\mur^2)  \nn
\\ 
&=& \int d\Phi_3 \, {\mathrm{Tr}}\left[ \mathbf{H} (Q^2, \Phi_3, \{m^2\},\muf^2, \mur^2)\; \mathbf{S} (N+1, Q^2, \Phi_3, \{m^2\},\mur^2)  \right] \nonumber \\
&\times&\,\Delta^i(N+1, Q^2,\muf^2, \mur^2 ) \Delta^j(N+1, Q^2,\muf^2, \mur^2 )  \;,
\end{eqnarray}
where the functions $\mathbf{H}, \mathbf{S}, \Delta$ are the hard, soft and initial state radiation factors, correspondingly. The first two, $\mathbf{H}$ and $\mathbf{S}$,  are matrices in colour space and depend on the three particle phase space, $\Phi_3$. The function $\mathbf{H}$ contains information on the hard-off shell dynamics and includes LO contributions as well as virtual corrections split into colour channels. The soft function, $\mathbf{S}$, describes the soft-wide angle emissions and is given by a solution of the relevant renormalization group equation. The initial state emissions, $\Delta^i\Delta^j$ take into account soft/collinear emissions from the initial state partons $i$ and $j$. The cross sections
depend on the renormalisation scale $\mur$, the factorisation scale $\muf$, and the masses of particles (squared) $\{m ^2\}$.

The inclusive total cross section is computed by integrating the expression over $Q^2$. For the differential distributions of an observable $\cal O$, in addition to the integration over $Q^2$, a function $\cal{F}_{\cal{O}}$ is introduced which includes a phase space restriction defining the observable $\cal O$ at the hand of a delta function:
\begin{eqnarray}
\label{eq:res:fact_diff}
\frac{d\tilde \sigh^{{\rm (NNLL)}}_{ij\tosv kl B}}{d\cal{O}}&&\hspace{-0.9cm}(N, {\cal O},\{m^2\}, \muf^2,\mur^2)  \nn
\\ 
&=& \int dQ^2\int d\Phi_3 \, {\mathrm{Tr}}\left[ \mathbf{H} (Q^2, \Phi_3, \{m^2\},\muf^2, \mur^2)\; \mathbf{S} (N+1, Q^2, \Phi_3, \{m^2\},\mur^2)  \right] \nonumber \\
&\times&\,\Delta^i(N+1, Q^2,\muf^2, \mur^2 ) \Delta^j(N+1, Q^2,\muf^2, \mur^2 ) \, {\cal F_{\cal{O}}}\left(Q^2, \Phi_3, \{m^2 \}\right)   \;.
\end{eqnarray}

The resummed hadronic $h_1 h_2$ cross sections of different accuracy denoted by ``res'' in the following are matched with the full NLO cross section according to
\bear
\label{hires}
\frac{d\sigma^{\rm (matched)}_{h_1 h_2 \tosv klB}}{d{\cal O}}({\cal O},\{m^2\},\muf^2, \mur^2) &=& 
\frac{d\sigma^{\rm (NLO)}_{h_1 h_2 \tosv kl B}}{d{\cal O}}({\cal O},\{m^2\},\muf^2, \mur^2) \\ \nn
&+&   \frac{d \sigma^{\rm
		(res-exp)}_
	{h_1 h_2 \tosv kl B}}{d{\cal O}}({\cal O},\{m^2\},\muf^2, \mur^2) 
\eear
with
\bear
\label{invmel}
&& \frac{d \sigma^{\rm
		(res-exp)}_{h_1 h_2 \tosv kl B}}{d{\cal O}} ({\cal O},\{m^2\},\muf^2, \mur^2) \! =   \sum_{i,j}\,
\int_{\sf C}\,\frac{dN}{2\pi
	i} \; \rho^{-N} f^{(N+1)} _{i/h{_1}} (\muf^2) \, f^{(N+1)} _{j/h_{2}} (\muf^2) \nn \\ 
&& \! \times\! \left[ 
\frac{d \tilde\sigh^{\rm (res)}_{ij\tosv kl B}}{d{\cal O}} (N,{\cal O},\{m^2\},\muf^2, \mur^2)
-  \frac{d \tilde\sigh^{\rm (res)}_{ij\tosv kl B}}{d{\cal O}} (N,{\cal O},\{m^2\},\muf^2, \mur^2)
{ \left. \right|}_{\scriptscriptstyle({\rm NLO})}\, \! \right], 
\eear
where ``res'' can refer to either NLL or NNLL accuracy. Correspondingly, ``matched'' corresponds to NLO + N(N)LL predictions, i.e.\ N(N)LL resummed results matched to NLO, either only to NLO in QCD or to complete NLO QCD and EW corrections. Additionally, we include results at the accuracy referred to as ``NLLwC''\footnote{NLLwC is also referred to as NLL$^\prime$ in the literature.}, where the $N$ independent ${\cal O}(\als)$ contributions in the expansion of the hard and soft functions, formally giving terms beyond NLL accuracy, are also included, see ~\cite{Kulesza:2017ukk,Kulesza:2018tqz} for a detailed description. 
The moments of the parton distribution functions $f_{i/h}(x, \muf^2)$ are defined in the standard way 
$$
f^{(N)}_{i/h} (\muf^2) \equiv \int_0^1 dx \, x^{N-1} f_{i/h}(x, \muf^2)\,,
$$
and $ d \sigh^{\rm
	(res)}_{ij\tosv kl B}/ dQ^2 \left. \right|_{\scriptscriptstyle({\rm NLO})}$ represents the perturbative expansion of the resummed cross section truncated at NLO. The inverse Mellin transform (\ref{invmel}) is evaluated numerically using 
a contour ${\sf C}$ in the complex-$N$ space according to the ``Minimal Prescription'' method proposed in Ref.~\cite{Catani:1996yz}.

We include the electroweak effects additively while matching the resummed QCD calculation to the differential cross sections calculated at the complete NLO QCD and EW accuracy~\cite{Frederix:2018nkq}, from now indicated by NLO (QCD+EW). More specifically, this means that at the LO accuracy, apart from the  ${\cal O}(\als^2 \alpha)$ contributions, also the ${\cal O}(\als \alpha^2)$ and ${\cal O}(\alpha^3)$ terms are included. The complete NLO(QCD+EW) result, besides the ${\cal O}(\als^3 \alpha)$ correction, contains also the ${\cal O}(\als^2 \alpha^2)$, ${\cal O}(\als \alpha^3)$ and ${\cal O}(\alpha^4)$ corrections as well as the above-mentioned LO terms.
Thus the EW effects are included in our final predictions up to NLO in the fine structure constant $\alpha$. It has been shown~\cite{Broggio:2019ewu} that the differences between additive and multiplicative matching is small with the exception of extreme tails of the differential distributions, where the electroweak Sudakov and soft gluon logarithms both play a role. As we do not show predictions within the range where these effects become significant, the method of matching carries little relevance in the following. 

Since we disentangle the pdfs from the partonic cross section by means of the Mellin transformation, there is no access to the individual $x_1$ and $x_2$ fractions of momenta of incoming partons. Therefore the formalism is restricted to observables invariant under boosts from the hadronic center-of-mass frame to the partonic center-of-mass frame, or correspondingly observables in the partonic frame. For example this means the resummation cannot be performed for a single particle rapidity distribution, but instead it can be performed for two particle rapidity difference distributions.

It is important to note that the threshold variable is not adapted for each individual observable. Therefore the threshold logarithms themselves do not depend on any observable other than the invariant mass of the three final state particles. The dependence on the other kinematic observables enters in the LO cross section, virtual corrections and the soft function. For the case of the $2 \to 2 $ process of heavy quark production, there exists an alternative formulation of threshold resummation using the 1 particle inclusive (1PI) kinematics~\cite{Laenen:1998qw} instead of the pair-invariant mass (PIM) kinematics, an extension of which to triple particle production we are using here. The 1PI kinematic is naturally more suitable to describe certain observables, such as transverse momentum of one of the produced heavy quarks.  However, while the analytical expressions for resummed distributions are known, the numerical calculations of the all-order resummed cross sections have not yet been achieved  in direct QCD, even for $2 \to 2$ processes.  In the next section we check how well  our expanded resummed results approximate the NLO distributions. The checks make us confident that the three-particle invariant-mass resummation formalism is sufficient for these observables.

\section{Numerical results}
\label{s:results}

In this section we discuss in detail numerical results obtained for the total cross sections and differential distributions for the process $pp \to  \ttB$ ($B=H, Z, W$) at $\sqrt S=13$ TeV. Unless otherwise stated, all resummed results presented here include EW corrections implemented additively, i.e.\ through matching of the NLO(QCD+EW) to the N(N)LL result, as explained above, and are called NLO(QCD+EW)+NLL, NLO(QCD+EW)+NLLwC or \mbox{NLO(QCD+EW)+NNLL}, correspondingly. 

In order to estimate the sensitivity of the theoretical predictions to the choice of renormalization and factorization scales, and cross sections and distributions are calculated for different central scale choices $\mu_0$: 
\begin{align}
 \mu_0&=Q=\sqrt{\left( \sum_{i=t,\bar t,B} p_i \right)^2}  \,,\nn \\
 \mu_0&=\frac{Q}{2} \,, \nn \\
 \mu_0&=\frac{M}{2}=\frac{\sum_{i=t,\bar t,B}m_{i}}{2} \,, \nn \\
 \mu_0&=H_T=\sum_{i=t,\bar t,B} m_T(i) = \sum_{i=t,\bar t,B} \sqrt{m_{i}^2+p_T^2(i)} \,,\nn  \\
 \mu_0&=\frac{H_T}{2}.\nn 
\end{align}
The invariant mass $Q$ can be seen as a natural scale for the kinematics of the invariant mass threshold resummation, whereas $H_T$-related scales are a popular dynamical scale choice in calculations of differential quantities, see e.g.~\cite{hxswg}. Predictions for the total NLO cross sections are often provided in the literature for the fixed scale choice  $\mu_0=M/2$~\cite{Beenakker:2001rj, Reina:2001sf,Dittmaier:2011ti}. The results for the total cross sections and invariant mass distributions at NLO(QCD)+NNLL  for the scale choices $\mu_0=Q$, $\mu_0=M/2$ and the ``in between'' scale choice $\mu_0= Q/2$ were reported in~\cite{Kulesza:2017ukk, Kulesza:2017jqv} and in~\cite{Kulesza:2018tqz} for the $\tth$ and $\ttV$ production, respectively. We also presented first results for the $p_T$ distribution of the $Z$ bosons in the process $pp\to \ttZ$, including results for $\mu_0 = H_T/2$,  in~\cite{Kulesza:2019adl}.

The scale uncertainty for any prediction is estimated with the so-called 7-point method. All values are calculated for seven different pairs of $\muf$ and $\mur$ around the central scale $\mu_0$. The maximal and the minimal values of the seven combinations 
\begin{align}
 \left(\frac{\muf}{\mu_0},\frac{\mur}{\mu_0}\right)=(0.5,0.5),(0.5,1),(1,0.5),(1,1),(1,2),(2,1),(2,2) \nn
\end{align}
give then the uncertainty around the central value at $\left(\frac{\muf}{\mu_0},\frac{\mur}{\mu_0}\right)=(1,1)$.

In these numerical studies, we use the same input parameters as in the Higgs cross section working group (HXSWG) Yellow Report 4 \cite{hxswg} and our previous publication~\cite{Kulesza:2018tqz}:
 $m_H=125 \,\text{GeV},
 m_t=172.5 \,\text{GeV}, 
 m_W=80.385 \,\text{GeV},
 m_Z=91.188 \,\text{GeV},
 G_F=1.1663787 \cdot 10^{-5} \,\text{GeV}^{-2}$ .
The CKM matrix in the calculations of the $\ttW$ cross sections is taken diagonal, in accordance with the Yellow Report setup. Concerning parton distribution functions, we use the PDF4LHC15\_nlo\_30 set \cite{Butterworth:2015oua,Dulat:2015mca,Harland-Lang:2014zoa,Ball:2014uwa,Gao:2013bia,Carrazza:2015aoa} for NLO(QCD) cross sections and the NNLL results expanded up to ${\cal O}(\als^3)$. 
The NLO(QCD+EW) cross sections, as well as resummed results matched to them, are calculated with the LUXqed17\_plus\_PDF4LHC15\_nnlo\_100 set \cite{Butterworth:2015oua,Dulat:2015mca,Harland-Lang:2014zoa,Ball:2014uwa,Gao:2013bia,Carrazza:2015aoa,Manohar:2016nzj,Manohar:2017eqh}. The pdf error is only calculated for the NLO cross section since the resummation is not expected to influence the value of the pdf error in any significant way.

The squared LO(QCD) amplitudes for $t \bar t W^{+}/W^{-}/Z$ in the multiplet basis were calculated with {\tt FORM} \cite{Vermaseren:2000nd} and the colour package \cite{vanRitbergen:1998pn} and then cross checked with {\tt MadGraph5\_aMC@NLO} \cite{Alwall:2014hca} and {\tt PowHel} \cite{Garzelli:2011vp, Garzelli:2012bn}.
NLO cross sections were obtained with {\tt MadGraph5\_aMC@NLO} \cite{Alwall:2014hca,Frederix:2018nkq}. The QCD one loop virtual corrections needed for the hard colour matrix $\mathbf{H}^{(1)}$ were numerically extracted from {\tt PowHel} and {\tt MadGraph5\_aMC@NLO}.

All numerical results for resummed quantities were calculated  and cross-checked with two independent in-house Monte Carlo codes. We have checked that we reproduce the values for the NLO(QCD+EW) total cross sections quoted in~\cite{Frederix:2018nkq} and in~\cite{Broggio:2019ewu} in the corresponding setups. 

\subsection{Total cross sections}
\begin{figure}[h!]
\centering
\includegraphics[width=0.45\textwidth]{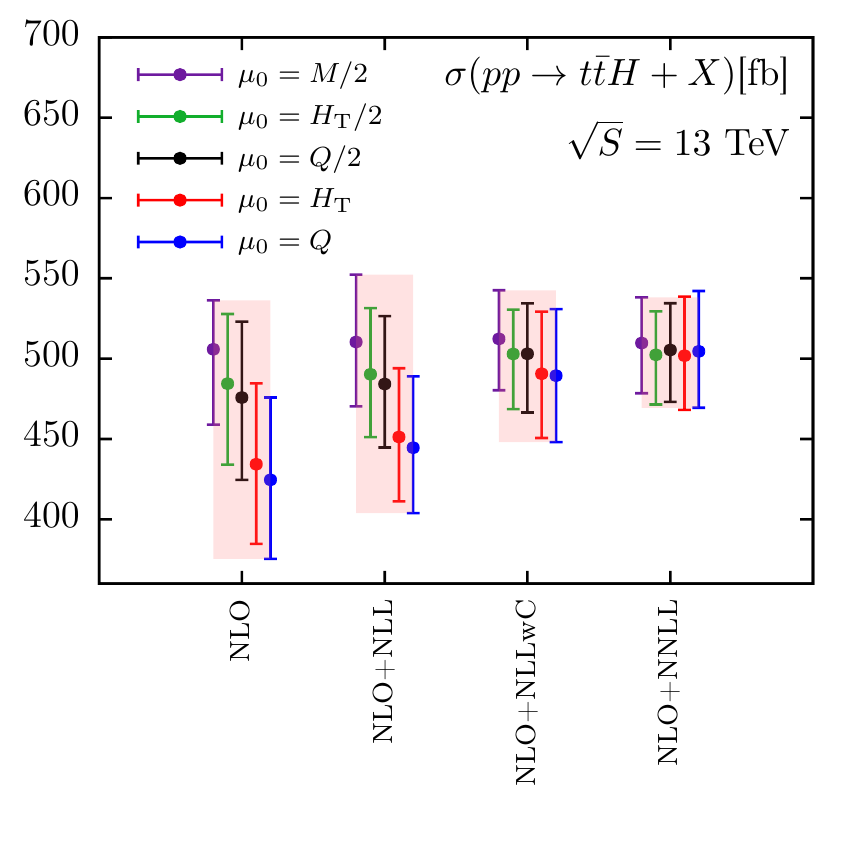}\\
\includegraphics[width=0.45\textwidth]{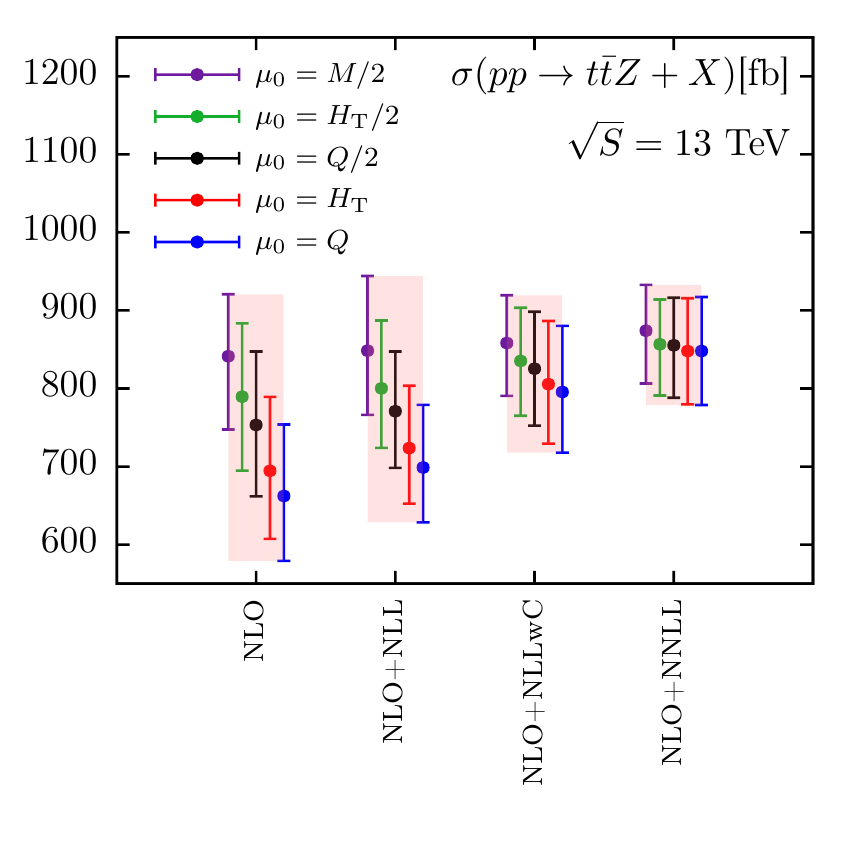}
\includegraphics[width=0.45\textwidth]{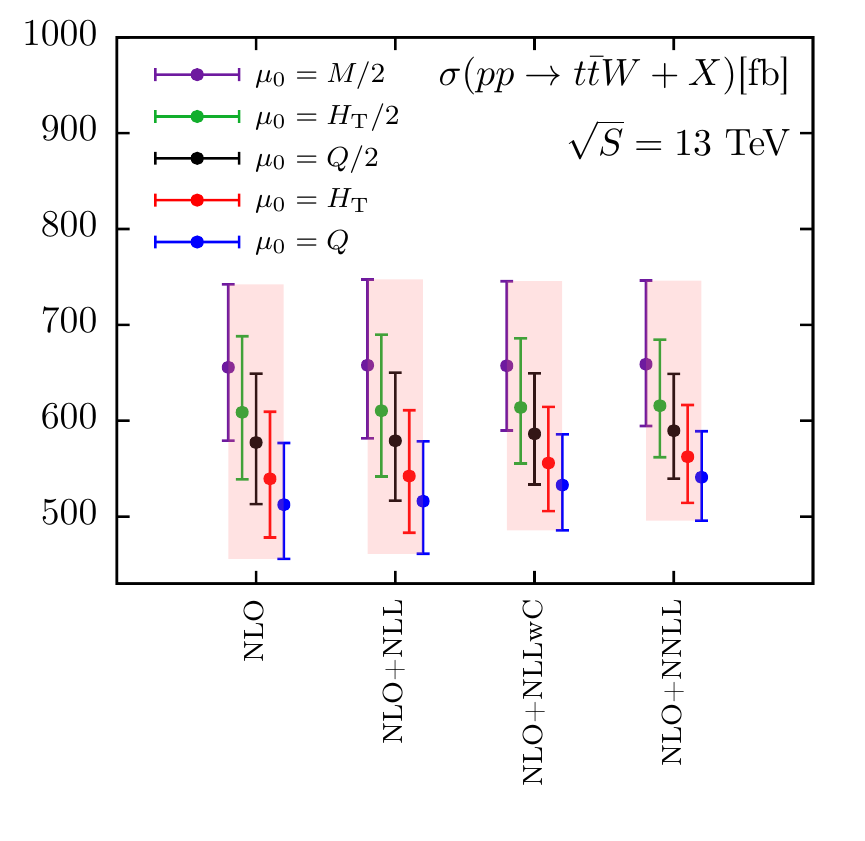}
\caption{Graphical illustration of the predictions in table~\ref{tab:totalxsec_EW_13TeV}. ``NLO'' stands here for NLO(QCD+EW).} 
\label{fig:totalxsec_13TeV_EW}
\end{figure}

We begin with the discussion of our results for the total cross sections for all three processes of associated top production with a heavy boson in $pp$ collisions at $\sqrt S=13$ TeV. They are listed in Table~\ref{tab:totalxsec_EW_13TeV} and graphically presented in Fig~\ref{fig:totalxsec_13TeV_EW}.  In order to keep the notation brief, we exceptionally refer to the NLO(QCD+EW) cross sections in Table~\ref{tab:totalxsec_EW_13TeV} and in Fig~\ref{fig:totalxsec_13TeV_EW} as ``NLO''.  For the predictions involving pure QCD corrections we refer the reader to~\cite{Kulesza:2017ukk, Kulesza:2019lye} and~\cite{Kulesza:2018tqz}. There we have also studied the quality of the approximation of the NLO total cross section provided by the $\cal O(\als)$ expansion of the resummed expression and concluded that  a big part of higher order corrections is indeed included in the resummed cross sections. Due to a similar impact of resummation on  the $\ttWp$ and $\ttWm$ cross sections,  for brevity we only show here results for their sum ($\ttW$).

\begin{table}[h!]
	\begin{center}
\renewcommand{\arraystretch}{1.5}
		\begin{tabular}{|c c c c c c c|}
			\hline
			process & $\mu_0$ & NLO{[}fb{]} & {NLO+NLL}{[}fb{]} & {NLO+NLLwC}{[}fb{]} & {NLO+NNLL}{[}fb{]} & $K_{\text{NNLL}}$ \tabularnewline
			\hline 
			$t \bar t H$ & $Q$ & $425_{-11.6\%}^{+12.1\%}$ & $445_{-9.2\%}^{+10.0\%}$  & $489_{-8.5\%}^{+8.4\%}$ &  $505_{-7.0\%}^{+7.5\%}$ & $1.19$ \tabularnewline
			& $H_T$ & $434_{-11.4\%}^{+11.6\%}$ & $451_{-8.9\%}^{+9.5\%}$  & $491_{-8.2\%}^{+7.9\%}$  & $502_{-6.7\%}^{+7.3\%}$ & $1.16$ \tabularnewline
			& $Q/2$ & $476_{-10.8\%}^{+9.9\%}$ & $484_{-8.2\%}^{+8.7\%}$  & $503_{-7.3\%}^{+6.2\%}$  & $505_{-6.4\%}^{+5.7\%}$ & $1.06$ \tabularnewline
			& $H_T/2$ & $484_{-10.4\%}^{+8.9\%}$ & $490_{-8\%}^{+8.4\%}$  & $503_{-6.8\%}^{+5.5\%}$  & $502_{-6.1\%}^{+5.4\%}$ & $1.04$ \tabularnewline
			& $M/2$ & $506_{-9.3\%}^{+6\%}$ & $510_{-7.8\%}^{+8.2\%}$  & $512_{-6.2\%}^{+5.9\%}$  & $510_{-6.1\%}^{+5.6\%}$ & $1.01$ \tabularnewline
			\hline 
			$t \bar t Z$  & $Q$ & $661_{-12.5\%}^{+13.8\%}$ & $698_{-10.1\%}^{+11.5\%}$  & $795_{-9.7\%}^{+10.6\%}$ &  $847_{-8.2\%}^{+8.1\%}$ & $1.28$ \tabularnewline
			& $H_T$ & $694_{-12.6\%}^{+13.6\%}$ & $723_{-9.8\%}^{+11.0\%}$  & $805_{-9.5\%}^{+10.0\%}$  & $848_{-8.0\%}^{+7.9\%}$ & $1.22$ \tabularnewline
			& $Q/2$ & $752_{-12.1\%}^{+12.5\%}$ & $770_{-9.4\%}^{+10.6\%}$  & $824_{-8.8\%}^{+8.8\%}$  & $854_{-7.8\%}^{+7.1\%}$ & $1.14$ \tabularnewline
			& $H_T/2$ & $788_{-11.9\%}^{+11.7\%}$ & $798_{-9.5\%}^{+10.7\%}$  & $834_{-8.4\%}^{+8.1\%}$  & $855_{-7.7\%}^{+6.6\%}$ & $1.09$ \tabularnewline
			& $M/2$ & $841_{-11.1\%}^{+9.4\%}$ & $848_{-9.7\%}^{+11.2\%}$  & $858_{-7.9\%}^{+7.1\%}$  & $874_{-7.8\%}^{+6.7\%}$ & $1.04$ \tabularnewline
			\hline
			$t \bar t W$ & $Q$ & $512_{-11.1\%}^{+12.5\%}$ & $516_{-10.6\%}^{+12.1\%}$  & $533_{-8.9\%}^{+9.9\%}$  & $541_{-8.4\%}^{+8.9\%}$ & $1.06$  \tabularnewline
			& $H_T$ & $539_{-11.3\%}^{+13.0\%}$ & $542_{-10.9\%}^{+12.6\%}$  & $556_{-9.0\%}^{+10.5\%}$  & $562_{-8.5\%}^{+9.6\%}$ & $1.04$ \tabularnewline
			& $Q/2$ & $577_{-11.1\%}^{+12.5\%}$ & $579_{-10.8\%}^{+12.3\%}$  & $586_{-9.0\%}^{+10.7\%}$  & $590_{-8.5\%}^{+10.0\%}$ & $1.02$ \tabularnewline
			& $H_T/2$ & $609_{-11.5\%}^{+13.0\%}$ & $610_{-11.2\%}^{+13\%}$  & $614_{-9.5\%}^{+11.8\%}$  & $616_{-8.8\%}^{+11.2\%}$ & $1.01$ \tabularnewline
			& $M/2$ & $656_{-11.7\%}^{+13.2\%}$ & $658_{-11.6\%}^{+13.6\%}$  & $657_{-10.3\%}^{+13.4\%}$  & $659_{-9.8\%}^{+13.3\%}$ & $1.00$ \tabularnewline	
			\hline 

		\end{tabular}
	\end{center}
\caption{Predictions for the total $pp \to t \bar t \, H/Z/W$ cross section at $\sqrt S=13$ TeV and different $\mu_0$. ``NLO'' stands here for NLO(QCD+EW). The listed error is the scale uncertainty calculated with the 7-point method.}
\label{tab:totalxsec_EW_13TeV}
\end{table}

Our NLO(QCD+EW)+NNLL results demonstrate remarkable stability w.r.t.\ different scale choices, delivering a compelling improvement of the theoretical predictions, specifically for processes involving a gluon channel. We see that the spread among the central values of the predictions, clearly visible at NLO(QCD+EW), is almost entirely eliminated for the $\ttZ$ and $\tth$ production processes. Moreover, for all scale choices, the scale uncertainty is reduced as the accuracy of the calculation improves. The degree of the improvement varies depending on the central scale, as well as the process,  reaching up to a factor of almost two. The effects are qualitatively similar, though less pronounced, for the $\ttW$ process.  

Comparing the NLO(QCD+EW)+NNLL cross sections obtained using  the \mbox{LUXqed17\_} \mbox{plus\_PDF4 LHC15\_nnlo\_100} pdf set with the NLO (QCD)+NNLL predictions obtained using the \mbox{PDF4LHC15\_nnlo\_30} set~\cite{Kulesza:2018tqz, Kulesza:2019lye} we observe that for the $\tth$  and the $\ttW$ production the EW effects lead to an increase (albeit very small, ca.\ 1\%, in the $\tth$ case) in the total cross sections, whereas the results for the $\ttZ$ production get only minimally affected and the differences are within the size of our statistical Monte Carlo uncertainty. This behaviour is inherited from the NLO(QCD+EW) and NLO(QCD) results, where in most cases the EW effects (obtained using a corresponding pdf set) lead to positive corrections, in agreement with~\cite{Frederix:2018nkq}.  
Since the EW corrections are introduced additively into the matched formula, c.f.\ Eq.~(\ref{invmel}), the effects of resummation are very similar to the pure QCD case. Correspondingly,  the NNLL K-factors, i.e.\ ratios of the NLO(QCD+EW)+NNLL total cross sections to the NLO(QCD+EW) total cross sections listed in Table~\ref{tab:totalxsec_EW_13TeV} and the ones obtained for the pure QCD cross sections quoted in~\cite{Kulesza:2017ukk},~\cite{Kulesza:2018tqz} and~\cite{Kulesza:2019lye} are very similar as the EW corrections impact them only minimally.

Given the observed improvement in stability of the predictions w.r.t.\ scale variation at NLO(QCD+EW)+NNLL, we  combine the predictions for our five scale choices, according to the envelope method proposed in~\cite{Dittmaier:2011ti}. The corresponding results are
\begin{align}
 \sigma_{t \bar t H}^{\rm NLO+NNLL}&=504 ^{+7.6 \% +2.4 \%}_{-7.1 \% -2.4 \%} \ {\rm fb} \,, \\
 \sigma_{t \bar t Z}^{\rm NLO+NNLL}&=859 ^{+8.6 \% +2.3 \%}_{-9.5 \% -2.3 \%} \ {\rm fb} \,, \\
 \sigma_{t \bar t W}^{\rm NLO+NNLL}&=592 ^{+26.1 \% +2.1 \%}_{-16.2 \% -2.1 \%} \ {\rm fb}
\end{align}
at $\sqrt{S}=13$\,TeV. 
The first error is the scale uncertainty while the second one is the PDF uncertainty of the NLO(QCD+EW) prediction.

As mentioned in the Introduction,  predictions for selected differential $t\bar t B$ distributions  at the NLO (QCD+EW) + NNLL accuracy were also calculated in the SCET approach \cite{Broggio:2019ewu}. In principle, a comparison of results obtained in the two different approaches not only can be used as an independent check of the two calculations, but also deliver information on the size of the effects which are formally below the considered level of precision. It has to be noted though that the two approaches involve two different sets of scales: our direct QCD calculations depend only on the factorization and renormalization scales, $\muf$ and $\mur$, while the SCET formalism involves the factorization, soft and hard scales, $\muf$, $\mu_S$ and $\mu_H$. Unfortunately, the treatment of the scales in the resummed and expanded parts as well as the scale choices made in~\cite{Broggio:2019ewu}: the $Q$-based $\{\muf=Q/2, \mu_H=Q, \mu_S=Q/\bar N\}$ and the $H_T$-based  $\{ \muf=H_T/2, \mu_H=H_T/2, \mu_S=H_T\bar N \}$ sets  cannot be directly translated into corresponding choices of  $\{ \muf , \mur \}$. Consequently,  no meaningful conclusions on the impact of subleading terms can be drawn by comparing only the central values. However, one can compare an overall behavior of the results as well as the values of the cross sections within their scale errors. We also need to point out that the error estimates are performed in different way. While we use the seven-point method, in Ref.~\cite{Broggio:2019ewu} each of the three scales, $\muf$, $\mu_H$ and $\mu_S$ is varied independently w.r.t.\ its central value by a factor of $2^{\pm 1}$, while the two other scales are frozen. The total uncertainty is then obtained by adding in quadrature the deviations corresponding to these three scales. Finally, as the most conservative estimate of the theoretical uncertainty, in the present study the envelope method is applied to five central scale choices: $\mu_0 = Q$, $H_T$, $Q/2$, $H_T/2$, $M/2$, in contrast to the two ($Q$-based and $H_T$-based) choices in Ref.~\cite{Broggio:2019ewu}. In all cases, the estimates of the pdf uncertainties agree at per mille point precision and will be not discussed further.

Using the scale envelope scale uncertainty, at NLO(QCD+EW)+NNLL we obtain 
$\sigma(\tth) =  504 ^{+7.6\%} _{-7.1\%}$~fb vs 
$\sigma(\tth) =  496 ^{+7.8\%} _{-5.9\%}$~fb in Ref.~\cite{Broggio:2019ewu}. The results agree well within the error bars and the theoretical uncertainties are also close in value to each other. Although the results for $\sigma(\ttZ)$ also agree within errors, we find here a bigger difference, i.e. $\sigma(\ttZ) = 859 ^{+8.6\%} _{-9.5\%}$~fb vs
$\sigma(\ttZ) = 811 ^{+11.0\%} _{-9.6\%}$ in Ref.~\cite{Broggio:2019ewu}. If we adjust the value of the top quark mass to the same value as used in Ref.~\cite{Broggio:2019ewu} ($m_t=173.34$~GeV), the envelope value of our result reduces to $846^ {+ 8.3\%}_{ - 9.5\%}$~fb. Nevertheless, the percentage difference between our results and that of Ref.~\cite{Broggio:2019ewu} is bigger for $\sigma(\ttZ)$ than for $\sigma(\tth)$. This can be traced back to the overall higher $K_{\rm NNLL}$ values for the $\ttZ$ process, as well as a bigger spread of  $K_{\rm NNLL}$ values in our approach. The latter leads to a much smaller spread of our NLO(QCD+EW)+NNLL predictions compared to  Ref.~\cite{Broggio:2019ewu}, corresponding to a difference in stability of the NLO(QCD+EW)+NNLL results between 1\% level for our calculations and 3\% for~\cite{Broggio:2019ewu}.

For the $\ttW$ production we obtain $\sigma(\ttW) = 592 ^{+26.1\%} _{-16.2\%}$~fb 
vs $\sigma(\ttW) = 582 ^{+13.4\%} _{-8.2\%}$~fb in Ref.~\cite{Broggio:2019ewu}, 
where we added the
predictions for $\ttW^+$ and $\ttW^-$. Here the central values are pretty close but we estimate the theoretical uncertainties to be significantly larger. This difference can be unambiguously explained by the wider central scale range used in our envelope. When the scale decreases from $Q$ to $M/2$, the NLO(QCD+EW) cross section grows by about 28\%. For a relatively smaller central scales span between $H_T/2$ and $Q/2$ used in ~\cite{Broggio:2019ewu}, the difference in the NLO(QCD+EW) cross section is only about 5.5\%. In both approaches, the impact of resummation is moderate, ranging from about 0.5\% at $\mu_0 = M/2$ to about 5\% at the highest scale $Q$ in our case and from 1 to 2 \% in~\cite{Broggio:2019ewu}. Consequently, the method of estimating the uncertainty of the NLO(QCD+EW) result leaves its imprint on the uncertainty of the NLO(QCD+EW)+NNLL predictions.  However, while we find that the resummation brings the central values of the predictions obtained with various scale choices closer together, Ref.~\cite{Broggio:2019ewu} reports that NLO(QCD+EW)+NNLL predictions for $Q$-based and $H_T$-based scale choices are more spread apart than  NLO(QCD+EW).

The main conclusion from the comparison of NLO(QCD+EW)+NNLL total cross sections  in Ref.~\cite{Broggio:2019ewu} and the present paper is that the results agree within the uncertainties. In most cases the estimated theoretical 
uncertainties are also similar, with one exception of $\ttW$ where there is a significant
difference of scale uncertainties due to different ranges of central scales taken into account
in the two approaches.

\subsection{Differential distributions}

\begin{figure}[h!]
\centering
\includegraphics[width=0.45\textwidth]{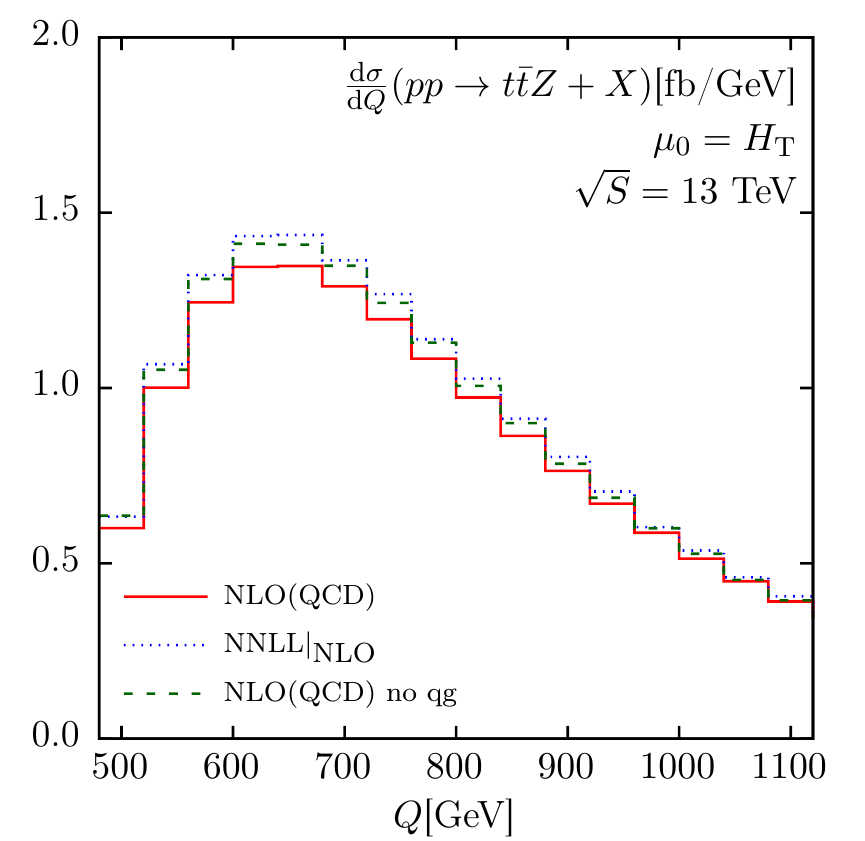}
\includegraphics[width=0.45\textwidth]{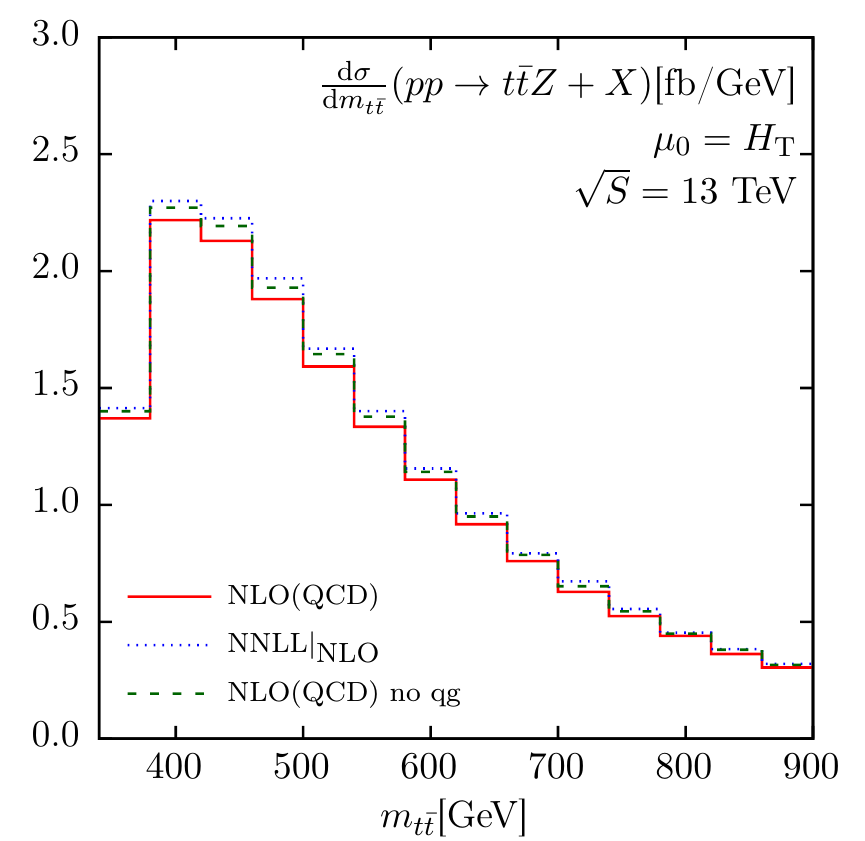}
\caption{Comparison between the expansion of the NNLL expression up to NLO accuracy in $\als$, the full NLO(QCD) result and the NLO(QCD) result without the $qg$ channels for the $pp \to \ttZ$ differential distributions in $Q$ and $m_{t \bar t}$.} 
\label{f:Qmttbardiff_NLOexp_ttZ}
\end{figure}

\begin{figure}[h!]
\centering
\includegraphics[width=0.45\textwidth]{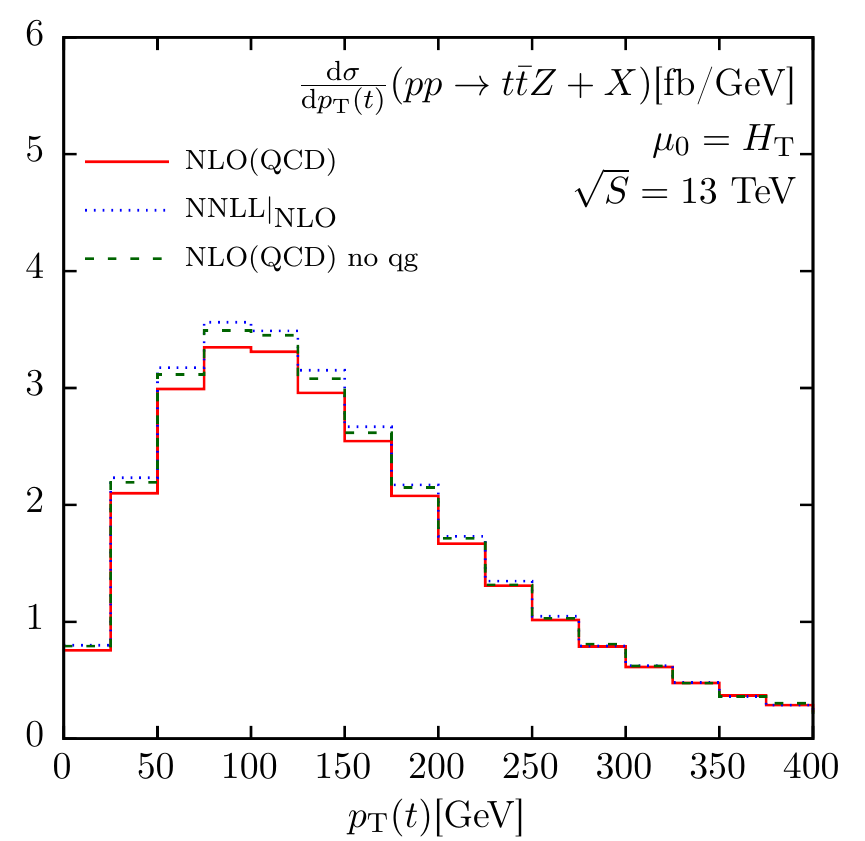}
\includegraphics[width=0.45\textwidth]{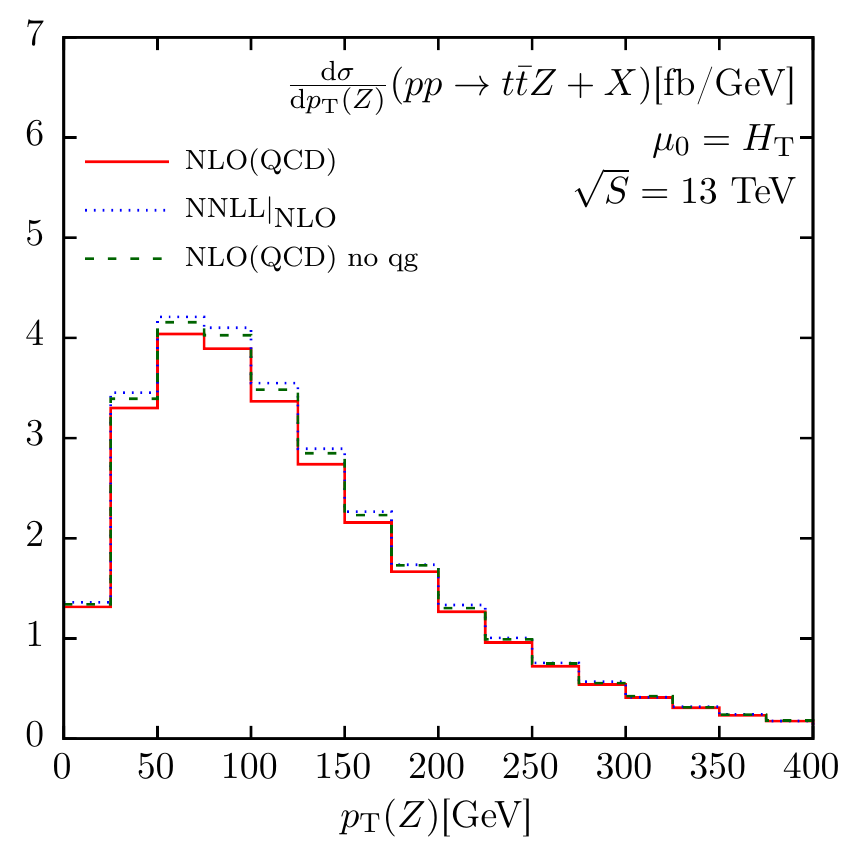}
\caption{The same as in Fig.~\ref{f:Qmttbardiff_NLOexp_ttZ} but for the $pp \to \ttZ$ differential distributions in $p_T(t)$ and $p_T(Z)$.} 
\label{f:pTtpTZ_NLOexp_ttZ}
\end{figure}

\begin{figure}[h!]
\centering
\includegraphics[width=0.45\textwidth]{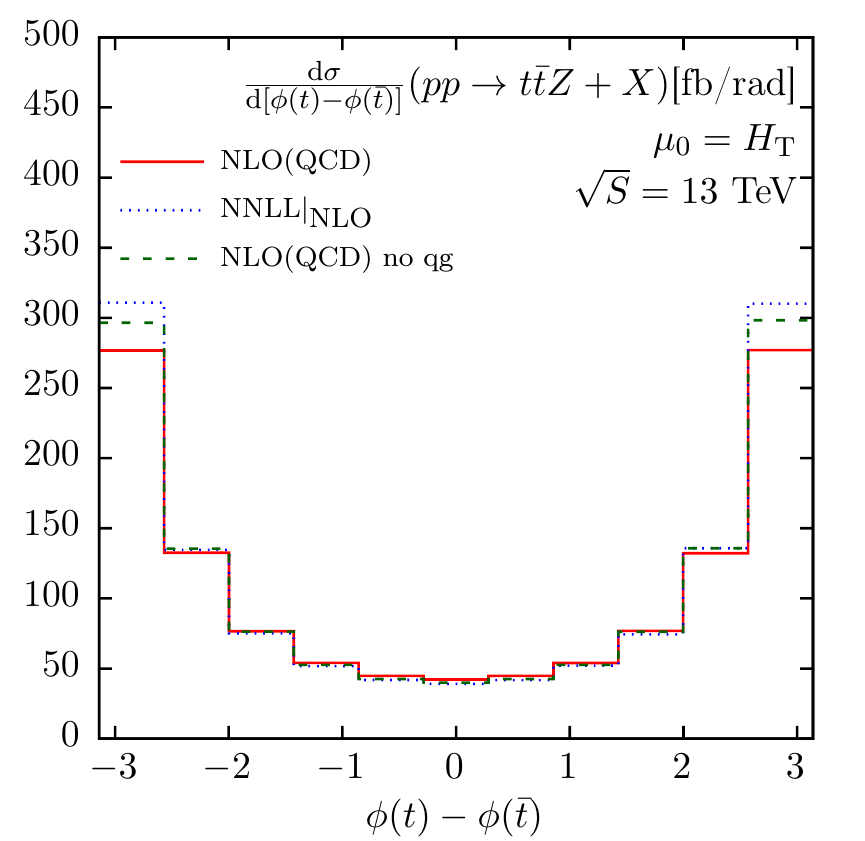}
\includegraphics[width=0.45\textwidth]{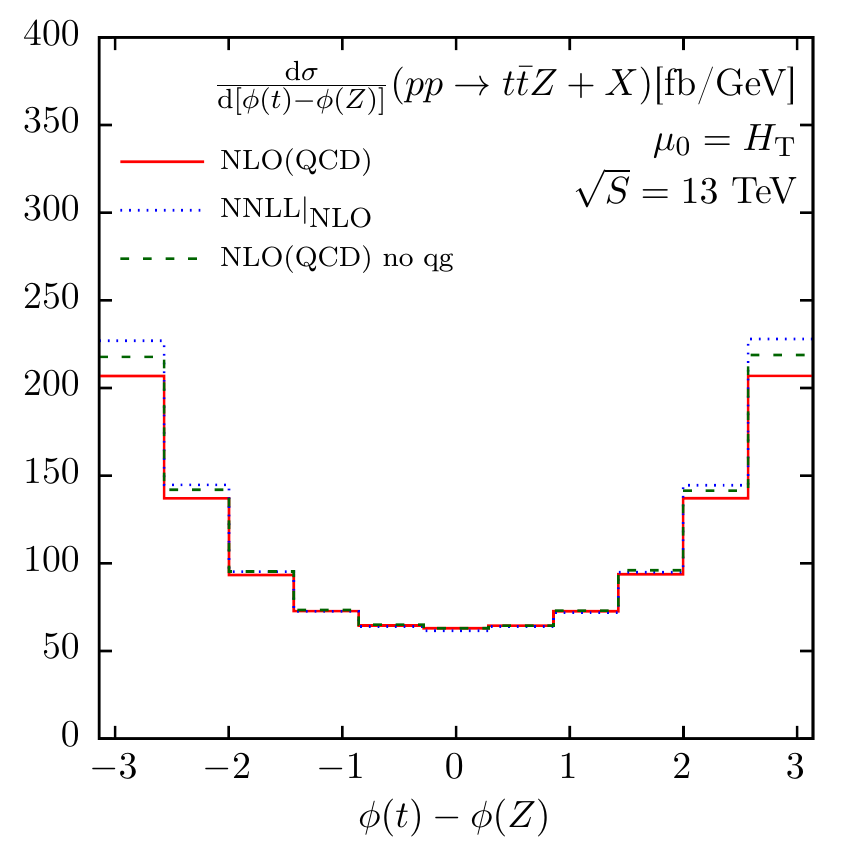}
\caption{The same as in Fig.~\ref{f:Qmttbardiff_NLOexp_ttZ} but for the $pp \to \ttZ$ differential distributions in $\phi(t) - \phi(\bar t)$ and $\phi(t) - \phi(Z)$.}
\label{f:phitminusphitbar_NLOexp_ttZ}
\end{figure}

\begin{figure}[h!]
\centering
\includegraphics[width=0.45\textwidth]{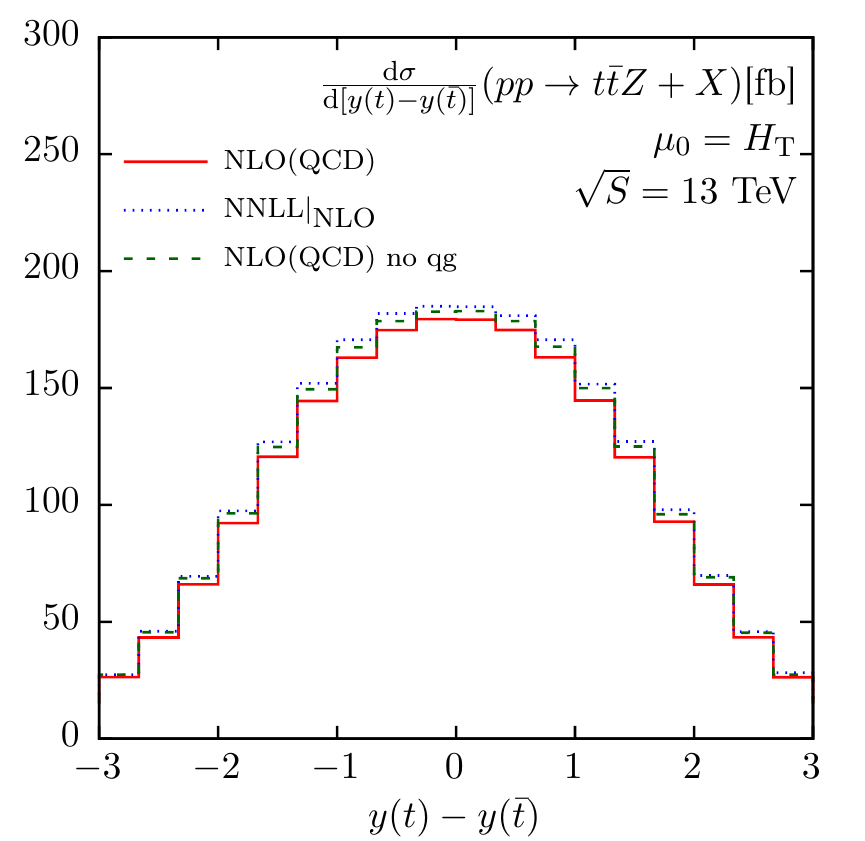}
\includegraphics[width=0.45\textwidth]{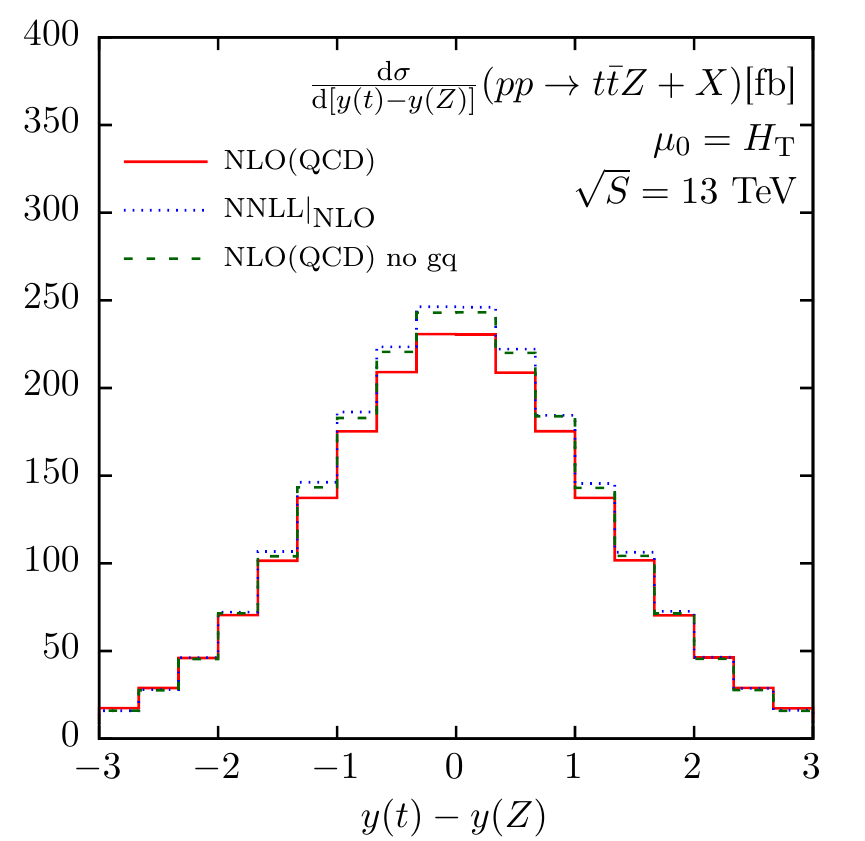}
\caption{The same as in Fig.~\ref{f:Qmttbardiff_NLOexp_ttZ} but for the $pp \to \ttZ$ differential distributions in $y(t)-y(\bar t)$ and $y(t)-y(Z)$. } 
\label{f:ytminusytbar_NLOexp_ttZ}
\end{figure}

In the following we present differential distributions for the processes $pp \to \ttB$ ($B=H,Z,W$). More specifically, these are distributions in the invariant mass $Q$ of the $t \bar t B$ system, the invariant mass $m_{t \bar t}$ of the $t \bar t$ pair, transverse momentum of the top quark $p_T (t)$, transverse momentum of the boson $p_T (B)$, the difference in rapidities between the top quark and the antitop quark $y(t) - y(\bar t)$, the difference in rapidities between the top quark  and the boson $y(t)-y(B)$, the difference in the azimuthal angle between the top quark and the antitop quark $\phi(t) - \phi(\bar t)$ and the difference in the azimuthal angle between the top quark and the boson $\phi(t) - \phi(B)$.

\begin{figure}[hb!]
\centering
\includegraphics[width=0.45\textwidth]{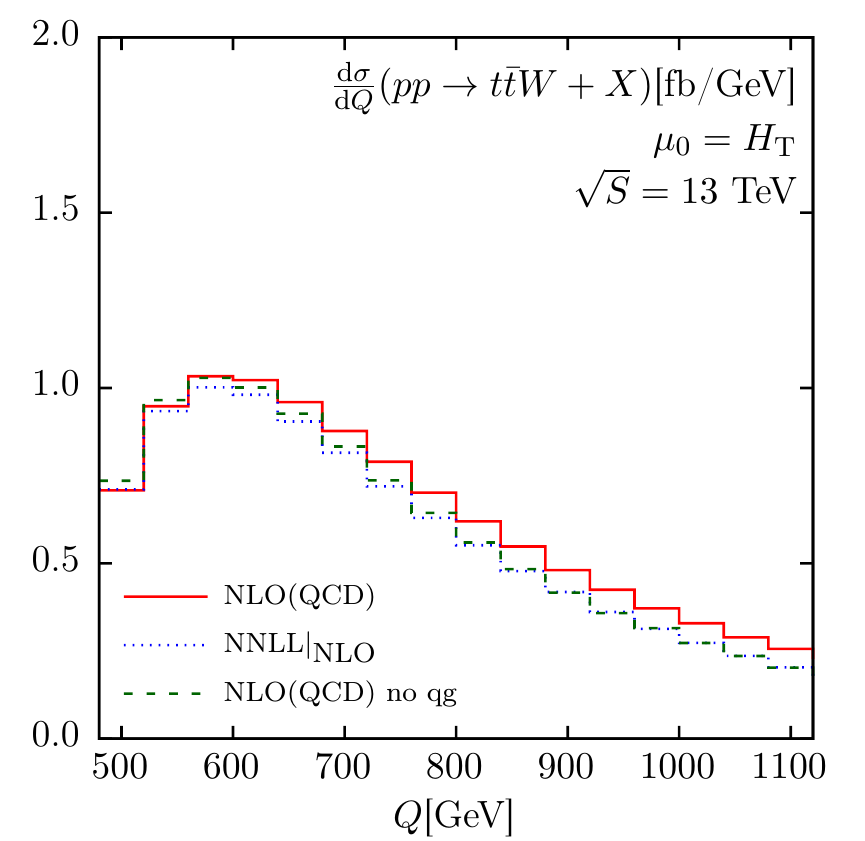}
\includegraphics[width=0.45\textwidth]{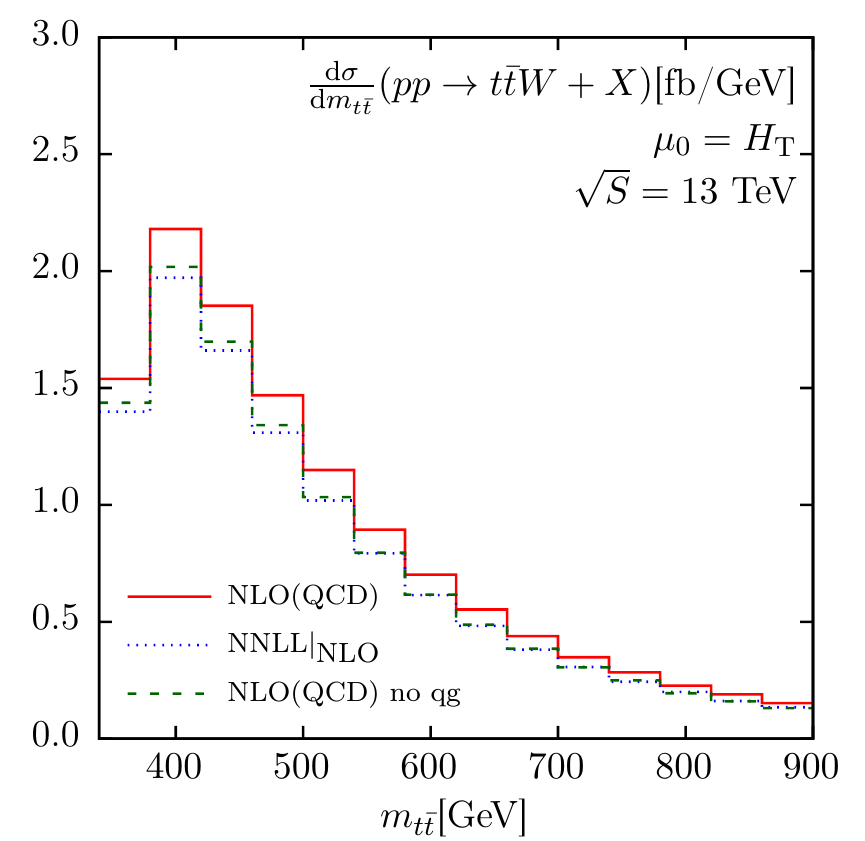}
\caption{Comparison between the expansion of the NNLL expression up to NLO accuracy in $\als$, the full NLO(QCD) result and the NLO(QCD) result without the $qg$ channels for the $pp \to \ttW$ differential distributions in $Q$ and $m_{t \bar t}$.} 
\label{f:Qmttbardiff_NLOexp_ttW}
\end{figure}

\begin{figure}[h!]
\centering
\includegraphics[width=0.45\textwidth]{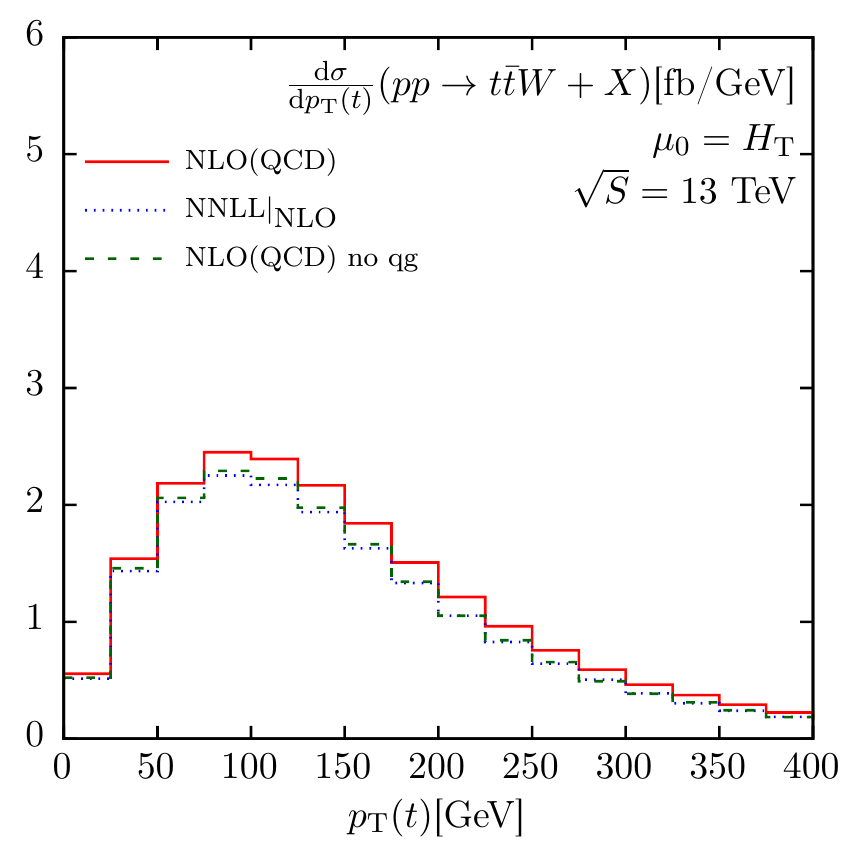}
\includegraphics[width=0.45\textwidth]{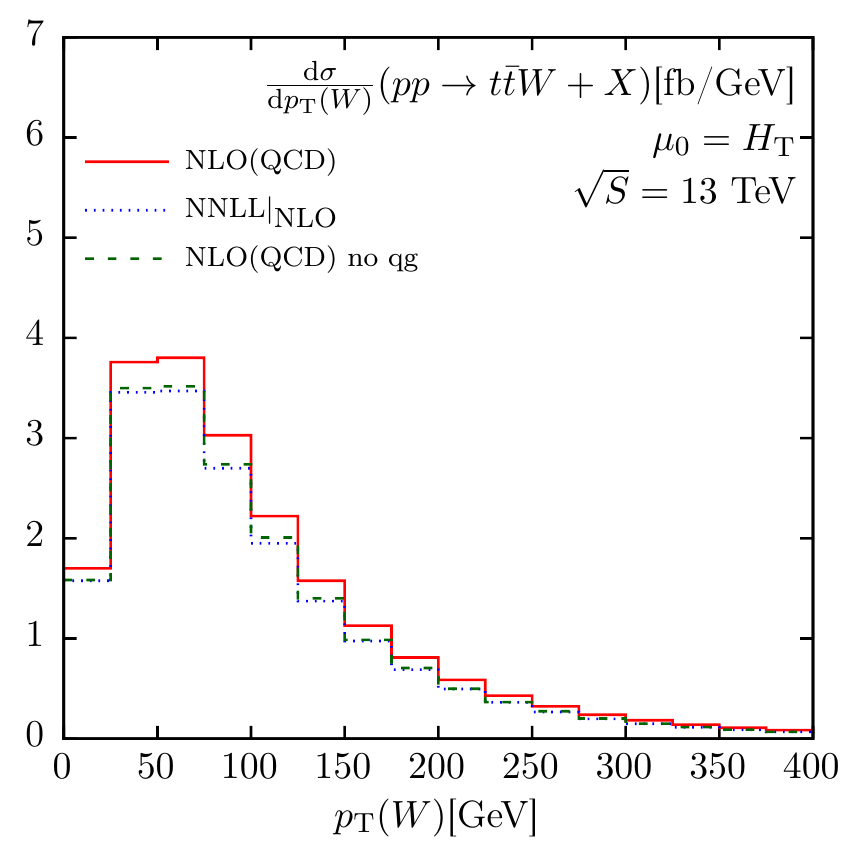}
\caption{The same as in Fig.~\ref{f:Qmttbardiff_NLOexp_ttW} but for the $pp \to \ttW$ differential distributions in $p_T(t)$ and $p_T(W)$.} 
\label{f:pTtpTZ_NLOexp_ttW}
\end{figure}

\begin{figure}[h!]
\centering
\includegraphics[width=0.45\textwidth]{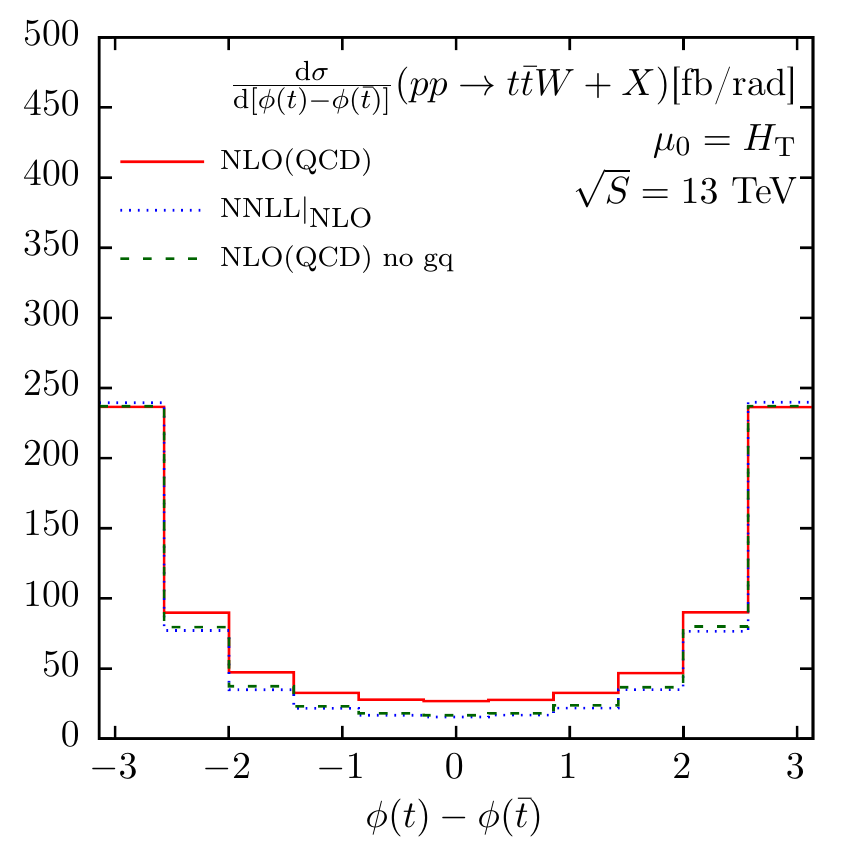}
\includegraphics[width=0.45\textwidth]{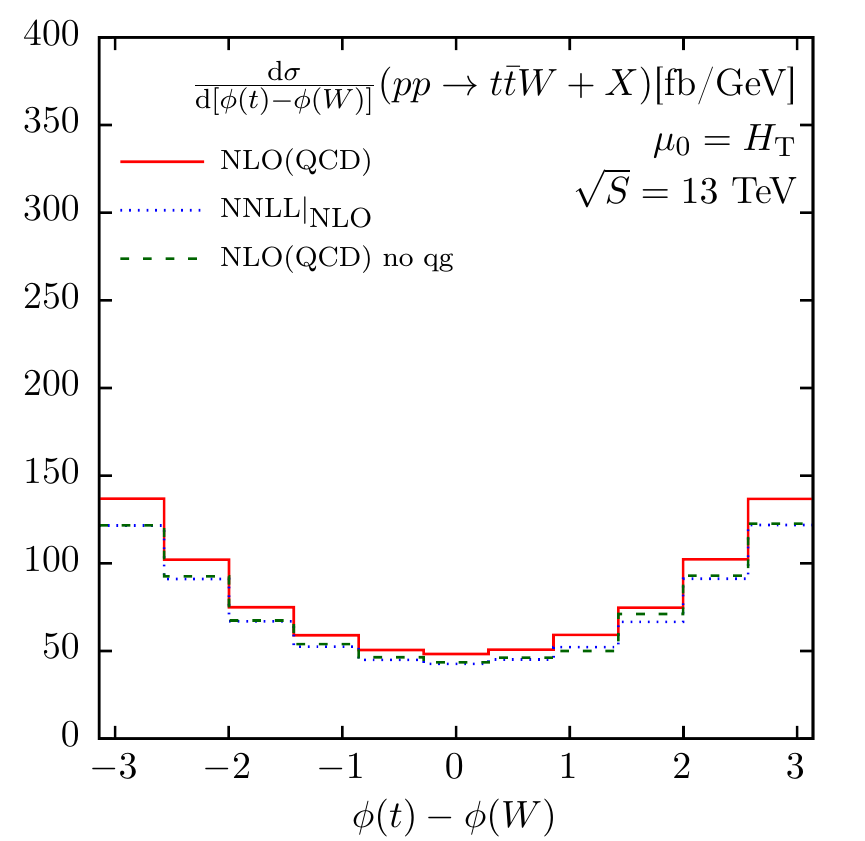}
\caption{The same as in Fig.~\ref{f:Qmttbardiff_NLOexp_ttW} but for the $pp \to \ttW$ differential distributions in $\phi(t) - \phi(\bar t)$ and $\phi(t) - \phi(W)$.}
\label{f:phitminusphitbar_NLOexp_ttW}
\end{figure}

\begin{figure}[h!]
\centering
\includegraphics[width=0.45\textwidth]{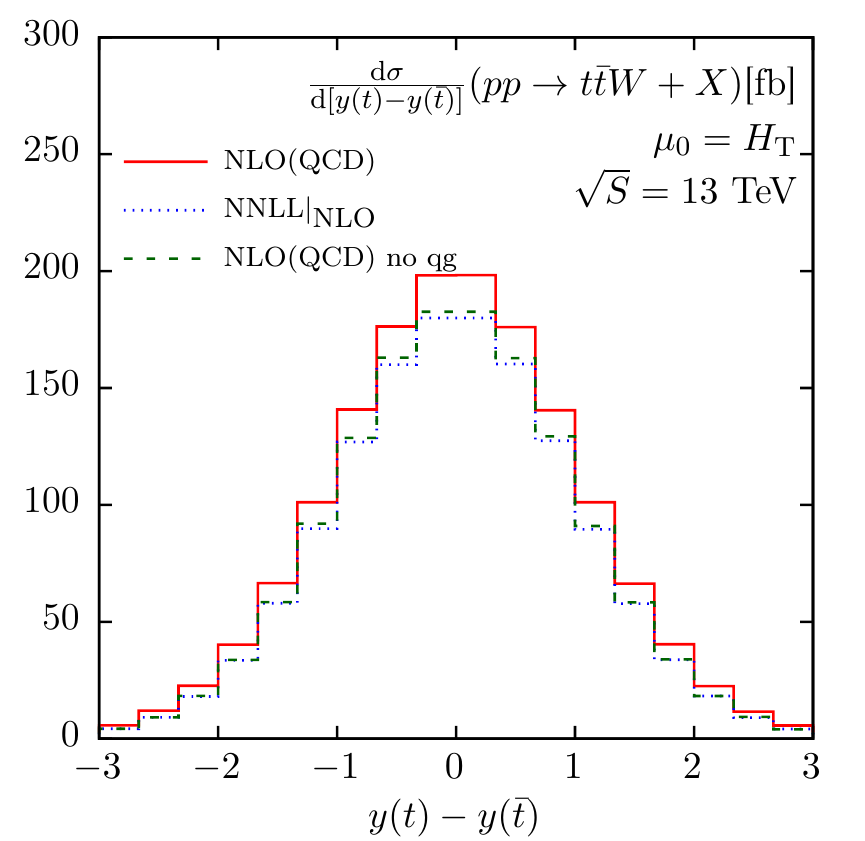}
\includegraphics[width=0.45\textwidth]{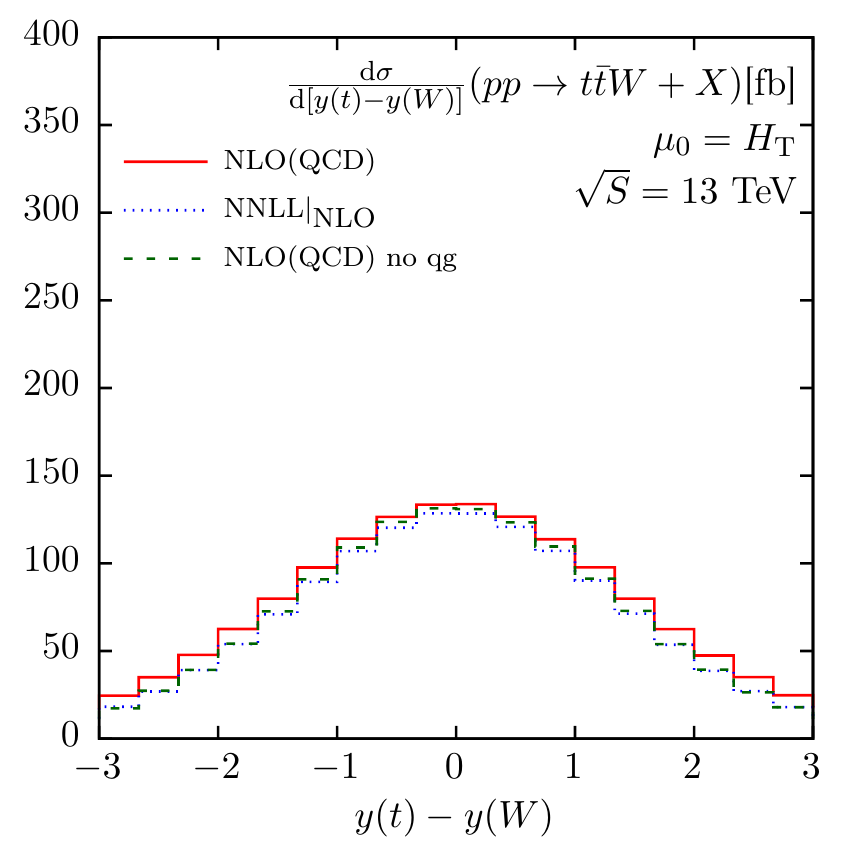}
\caption{The same as in Fig.~\ref{f:Qmttbardiff_NLOexp_ttW} but for the $pp \to \ttW$ differential distributions in $y(t)-y(\bar t)$ and $y(t)-y(W)$. } 
\label{f:ytminusytbar_NLOexp_ttW}
\end{figure}

In~\cite{Kulesza:2017ukk} and~\cite{Kulesza:2018tqz}, we showed that the resummed results for the total cross sections, expanded to the same order in $\als$ as that of NLO, approximates well the full NLO(QCD) cross section in the $q \bar q$ and $gg$ channels, especially for the $\tth$ and $\ttZ$ production. The $qg$ channel appears for the first time at NLO and no resummation is performed for this channel: it only enters the resummation-improved predictions via matching. Before commenting on individual differential distributions we first study if the statement regarding the quality of the approximation carries on to the differential level. Since the effects of higher order logarithmic corrections treated by resummation are very similar for the  $\tth$ and $\ttZ$ production and are very similar for these two processes, we present a corresponding study only for the $\ttZ$ process. We choose to analyze the quality of the approximation of the NLO(QCD) distributions by the expansion of the NNLL result at the scale $\mu_0=H_T$ where the resummation effects are very relevant. The results for all differential distributions mentioned above are shown in Figs.~\ref{f:Qmttbardiff_NLOexp_ttZ}, \ref{f:pTtpTZ_NLOexp_ttZ}, \ref{f:phitminusphitbar_NLOexp_ttZ} and \ref{f:ytminusytbar_NLOexp_ttZ}. Note that in this comparison the NLO(QCD) cross sections are used, as the judgement of the quality of the approximation concerns only the QCD corrections. We observe that the expanded NNLL differential distributions offer very good approximations
of the NLO(QCD) results for the distributions in $Q$, $m_{t \bar t}$, $p_T(t)$, $p_T(Z)$, $y(t)-y(\bar t)$ and $y(t)-y(Z)$. For the distributions in $\phi(t) - \phi(\bar t)$ and $\phi(t) - \phi(Z)$, the quality of the approximation is excellent for small angle differences. It worsens slightly with increasing angle but without exceeding 5\% difference in the largest angular difference bins. With this small exception, the demonstrated quality of the approximation let us conclude that the NNLL differential distributions for the $\ttZ$ production considered here take into account a big part of the higher order corrections to the $gg$ and $q \bar q$ channels. Next, we perform analogous studies for the process $pp \to \ttW$, see Figs.~\ref{f:Qmttbardiff_NLOexp_ttW}, \ref{f:pTtpTZ_NLOexp_ttW}, \ref{f:phitminusphitbar_NLOexp_ttW} and \ref{f:ytminusytbar_NLOexp_ttW}. Also in this case we observe that the expanded NNLL result provides a very good approximation of the NLO distributions when the $qg$ channel contributions are subtracted.

\begin{figure}[t!]
\centering
\includegraphics[width=0.48\textwidth]{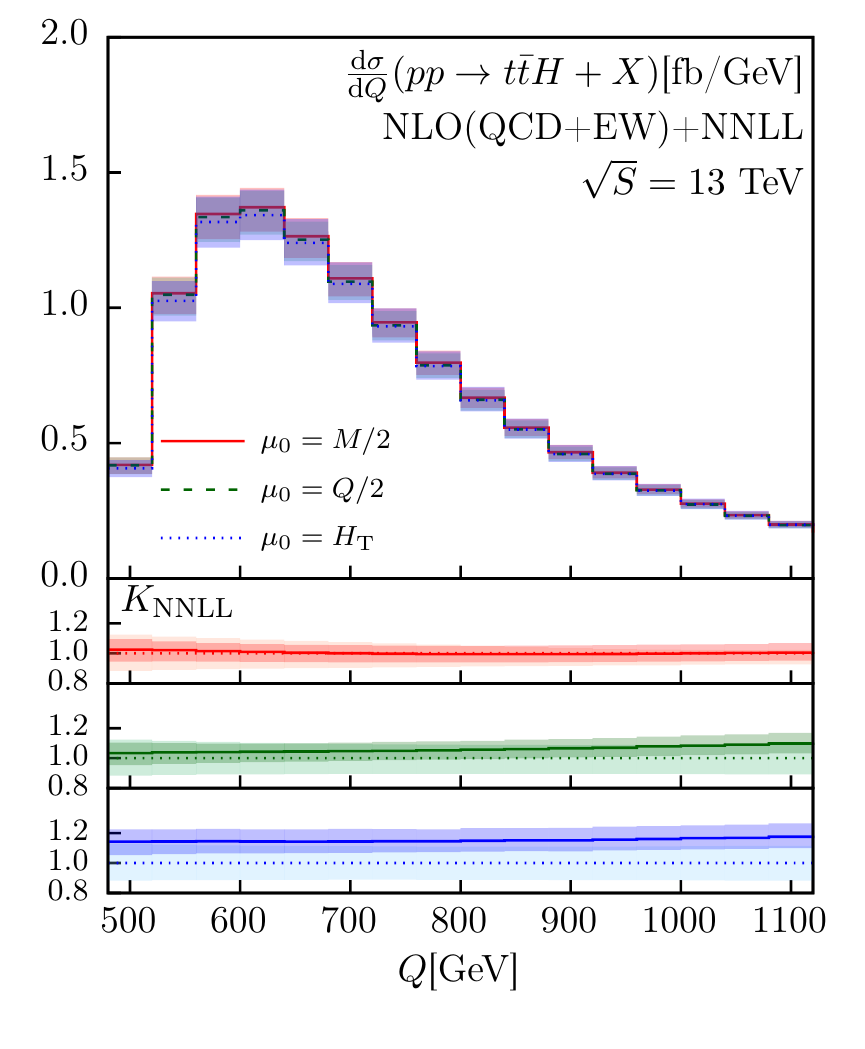}
\includegraphics[width=0.48\textwidth]{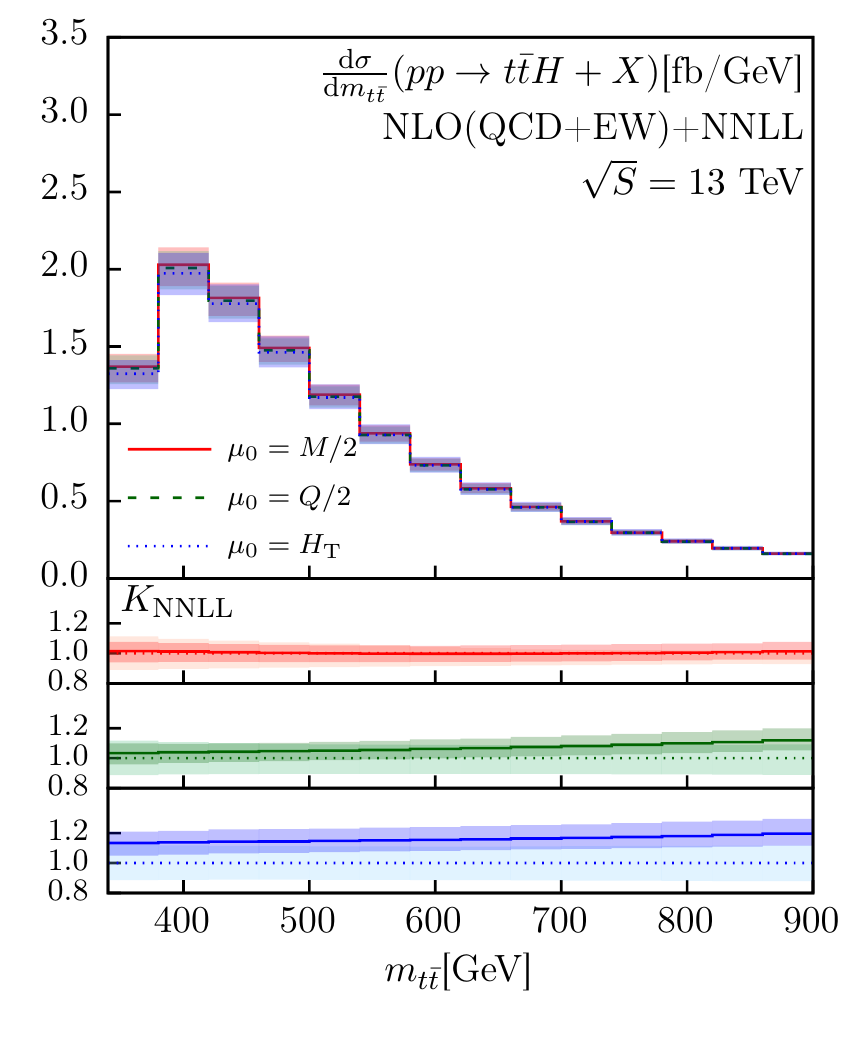}
\caption{Predictions for $pp \to \tth$ differential cross section in $Q$ and $m_{t \bar t}$. Lower panels show ratio of the NLO(QCD+EW)+NNLL and NLO(QCD+EW) distributions for three central scale choices $\mu_0=M/2$, $\mu_0=Q/2$ and $\mu_0=H_T$. Only scale uncertainties are shown.} 
\label{f:Qmttbardiff_ttH}
\end{figure}

\begin{figure}[h!]
\centering
\includegraphics[width=0.48\textwidth]{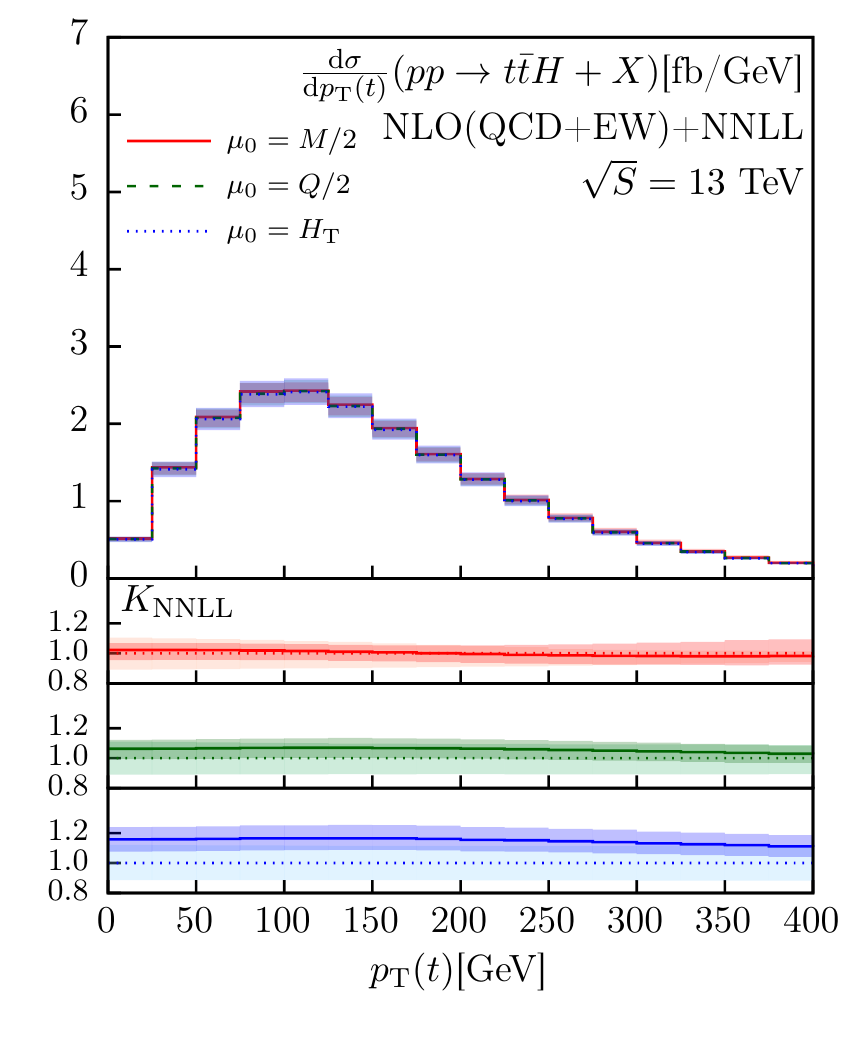}
\includegraphics[width=0.48\textwidth]{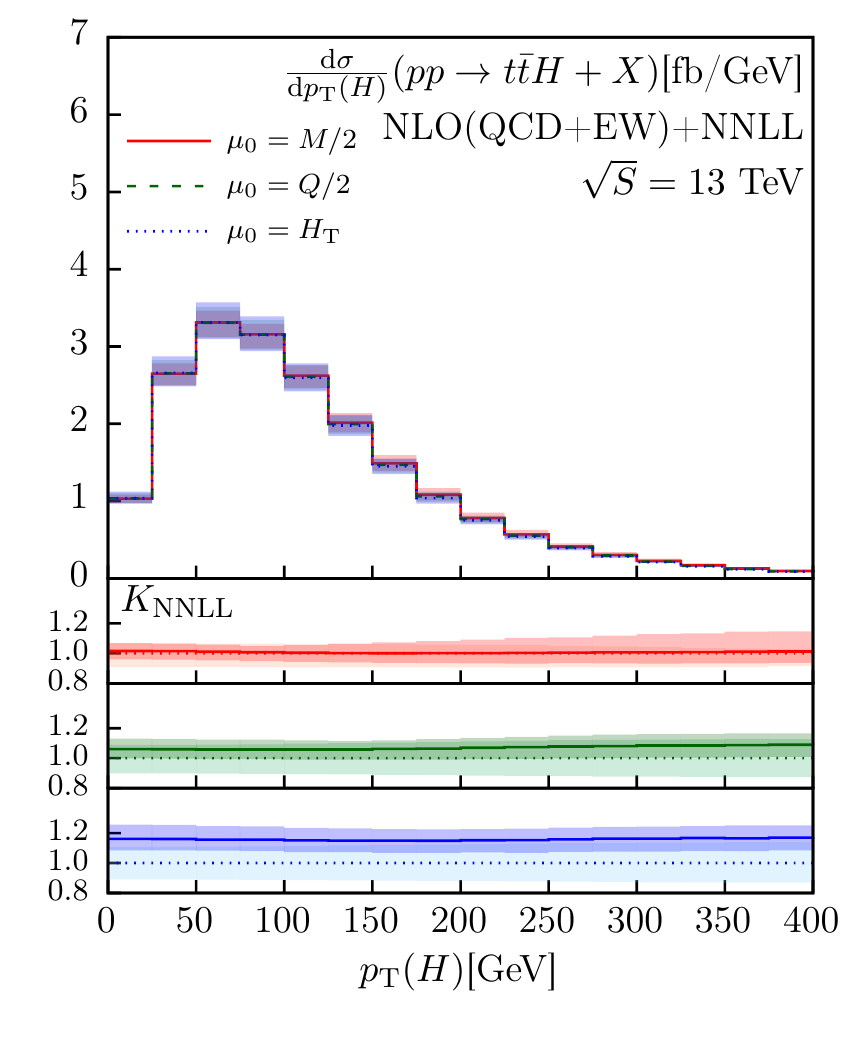}
\caption{Predictions for $pp \to \tth$ differential cross section in $p_T(t)$ and $p_T(H)$. Lower panels show ratio of the NLO(QCD+EW)+NNLL and NLO(QCD+EW) distributions for three central scale choices $\mu_0=M/2$, $\mu_0=Q/2$ and $\mu_0=H_T$. Only scale uncertainties are shown.} 
\label{f:pTdiff_ttH}
\end{figure}

\begin{figure}[h!]
\centering
\includegraphics[width=0.48\textwidth]{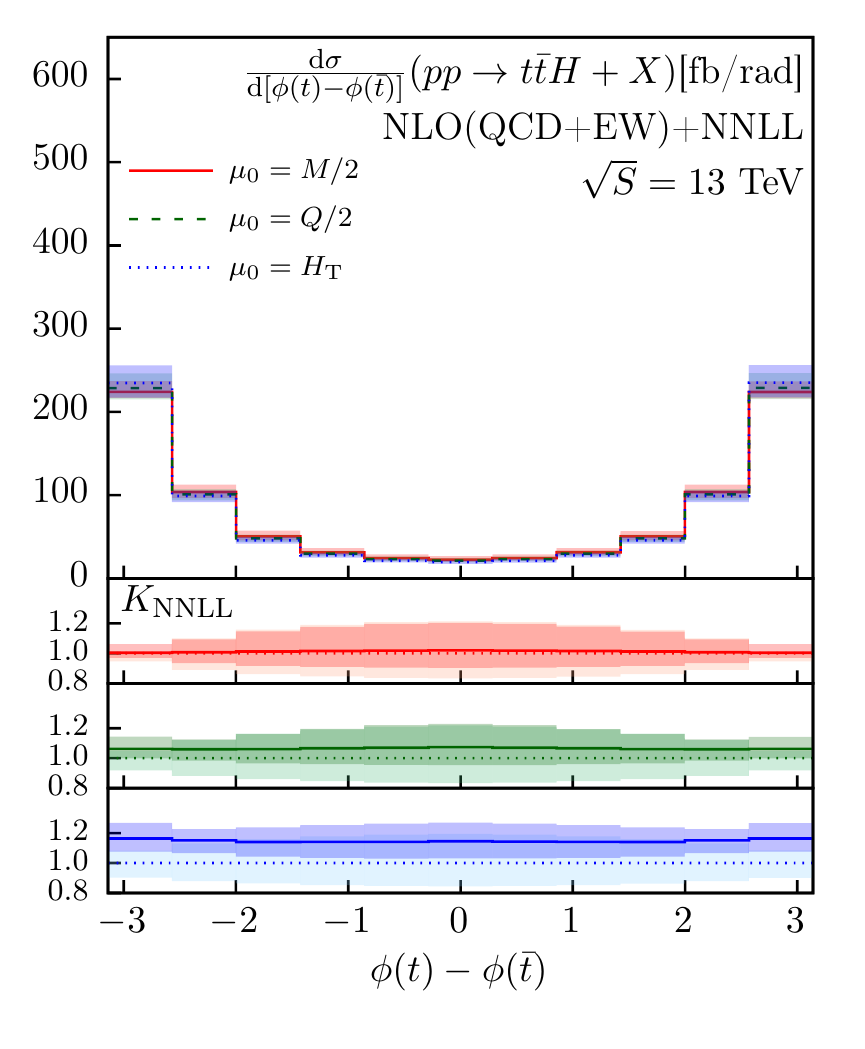}
\includegraphics[width=0.48\textwidth]{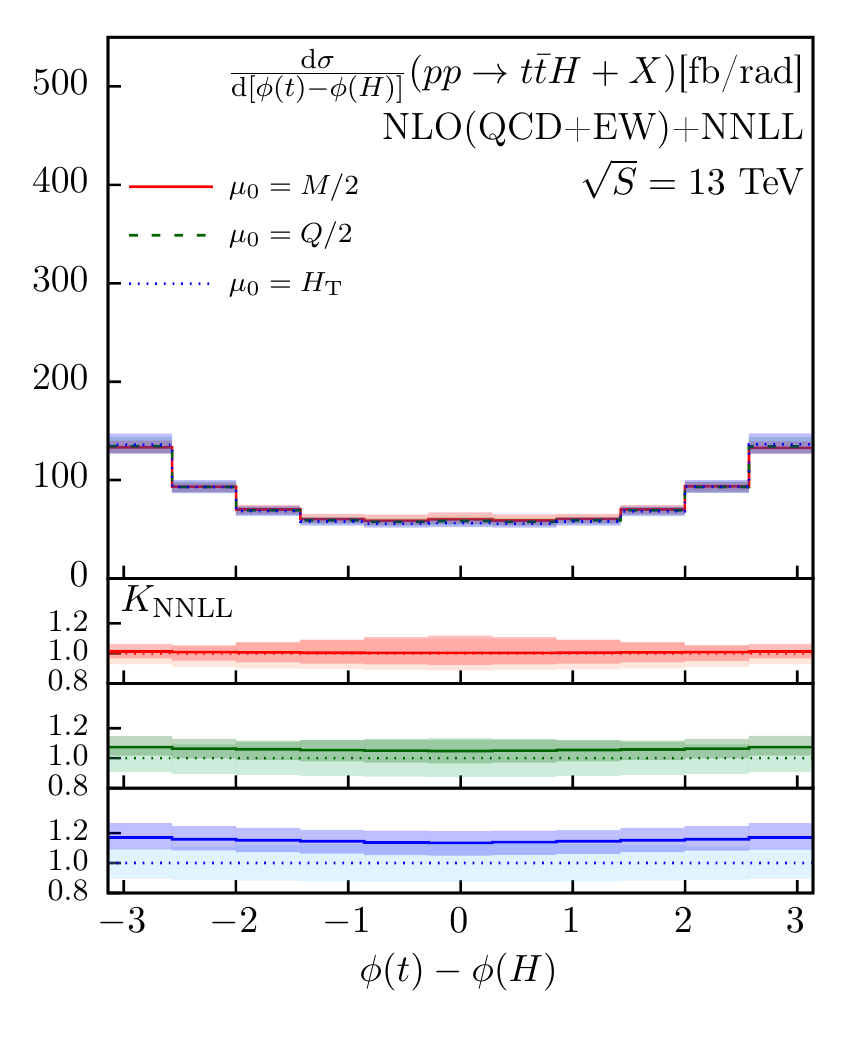}
\caption{Predictions for $pp \to \tth$ differential cross section in $\phi(t)-\phi(\bar t)$ and $\phi(t)-\phi(H)$. Lower panels show ratio of the NLO(QCD+EW)+NNLL and NLO(QCD+EW) distributions for three central scale choices $\mu_0=M/2$, $\mu_0=Q/2$ and $\mu_0=H_T$. Only scale uncertainties are shown.} 
\label{f:phidiff_ttH}
\end{figure}

\begin{figure}[h!]
\centering
\includegraphics[width=0.48\textwidth]{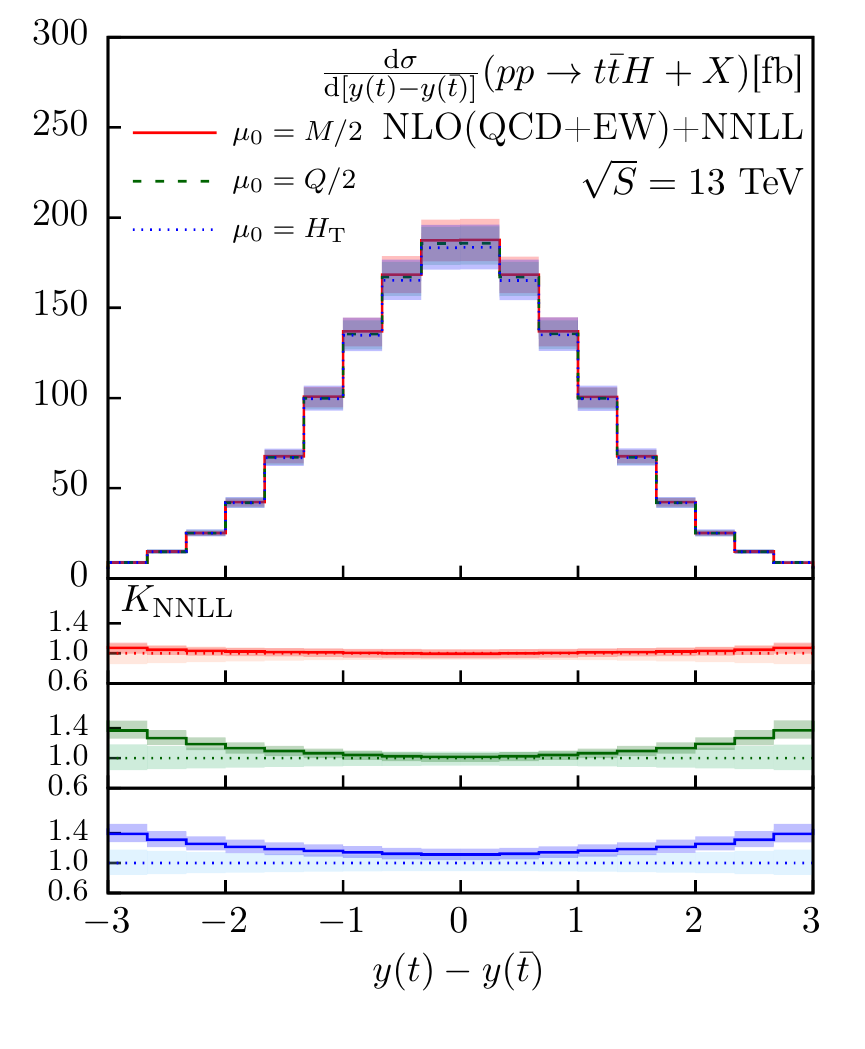}
\includegraphics[width=0.48\textwidth]{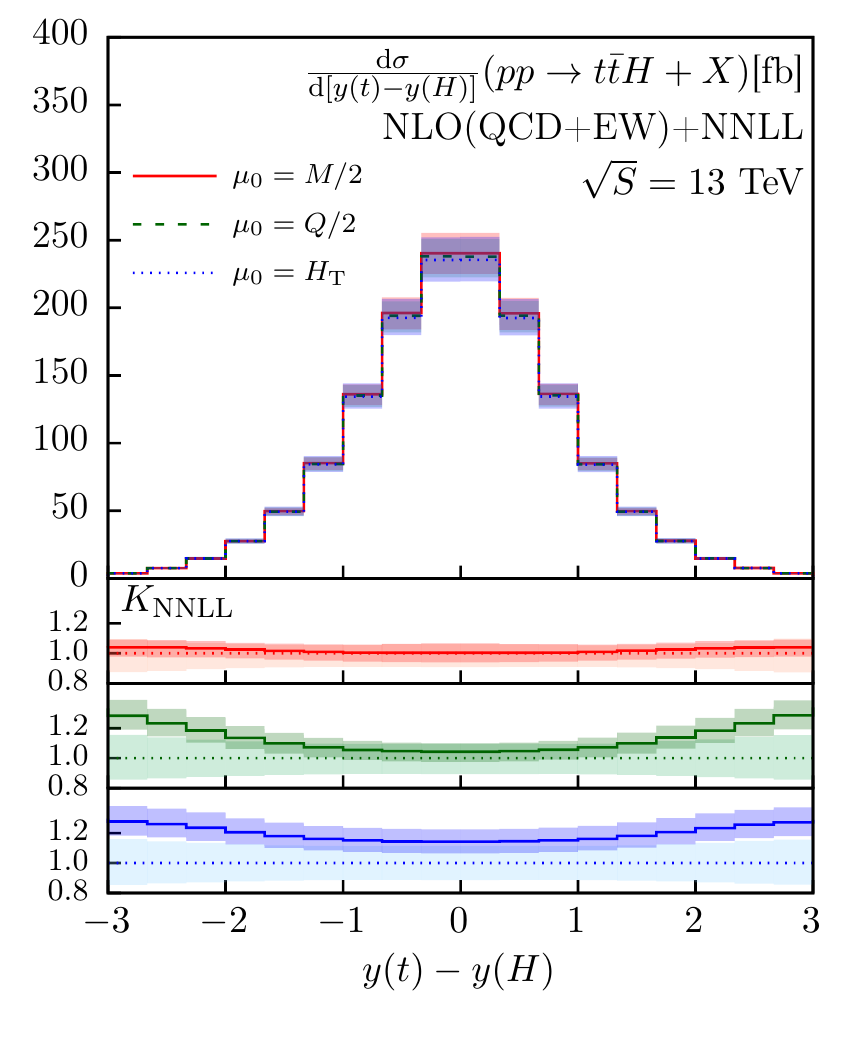}
\caption{Predictions for $pp \to \tth$ differential cross section in $y(t)-y(\bar t)$ and $y(t)-y(H)$. Lower panels show ratio of the NLO(QCD+EW)+NNLL and NLO(QCD+EW) distributions for three central scale choices $\mu_0=M/2$, $\mu_0=Q/2$ and $\mu_0=H_T$. Only scale uncertainties are shown.} 
\label{f:ydiff_ttH}
\end{figure}

\begin{figure}[h!]
\centering
\includegraphics[width=0.48\textwidth]{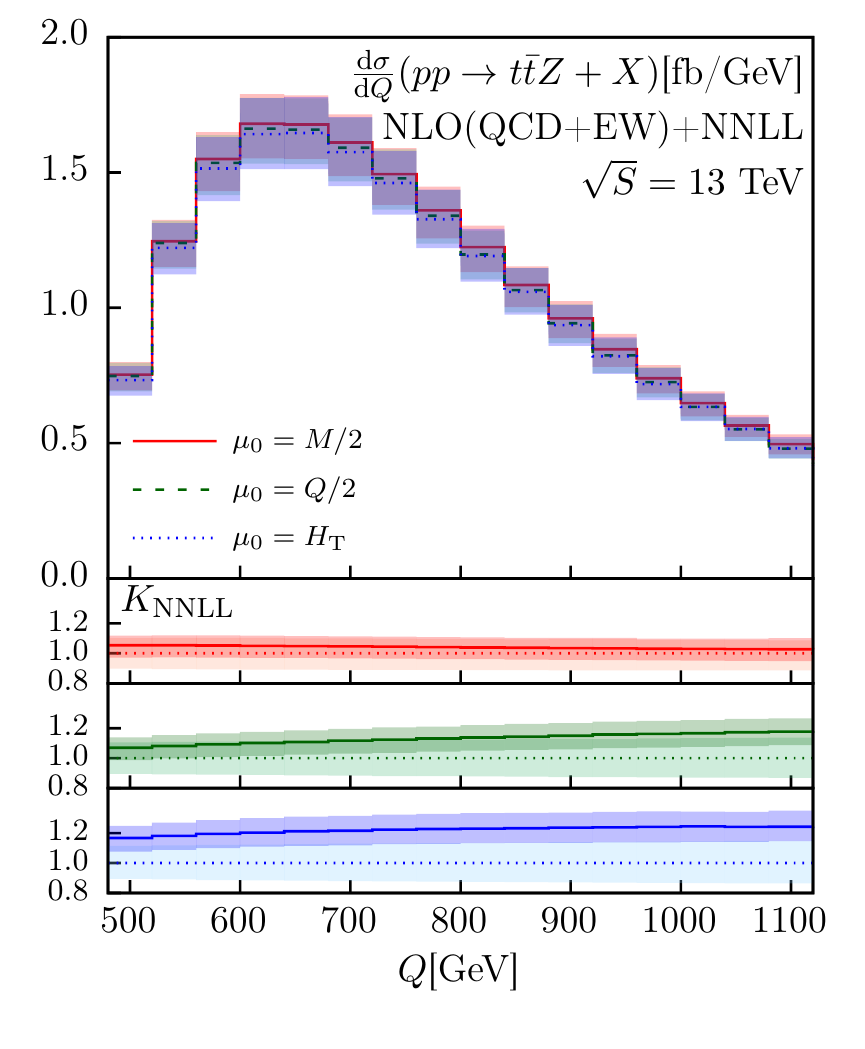}
\includegraphics[width=0.48\textwidth]{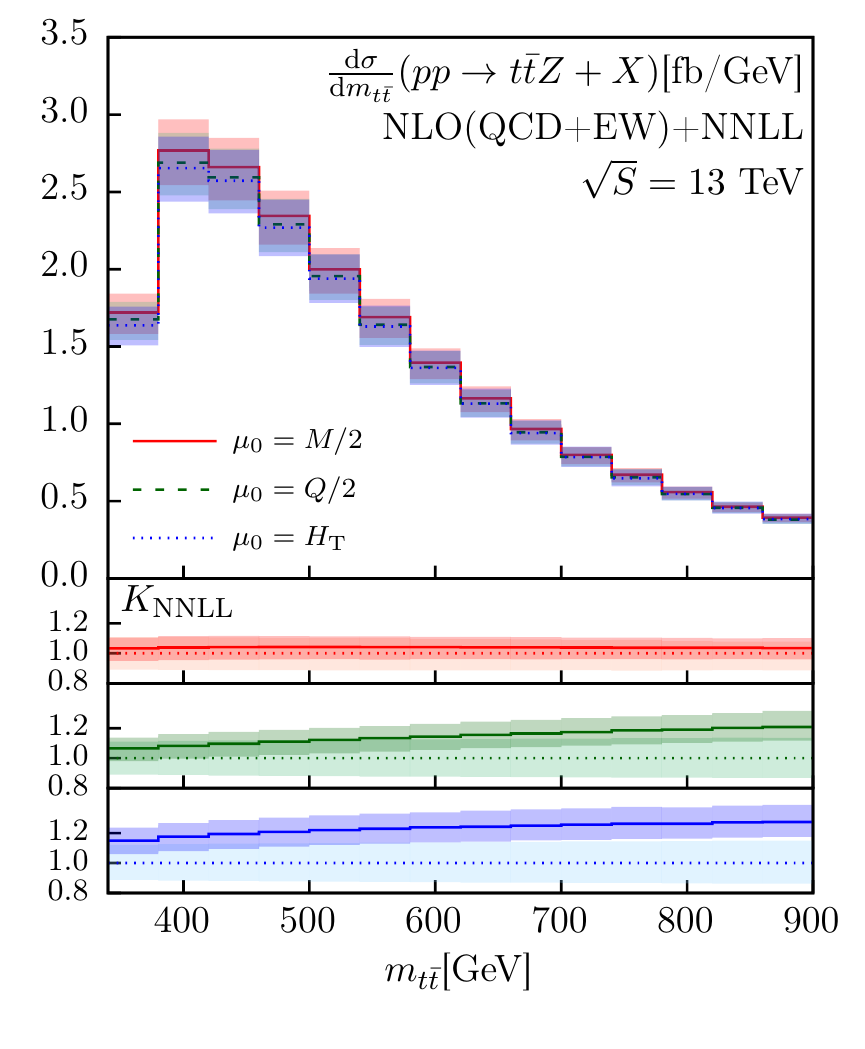}
\caption{The same as in Fig.~\ref{f:Qmttbardiff_ttH} but for the $pp \to \ttZ$ process.} 
\label{f:Qmttbardiff_ttZ}
\end{figure}

\begin{figure}[h!]
\centering
\includegraphics[width=0.48\textwidth]{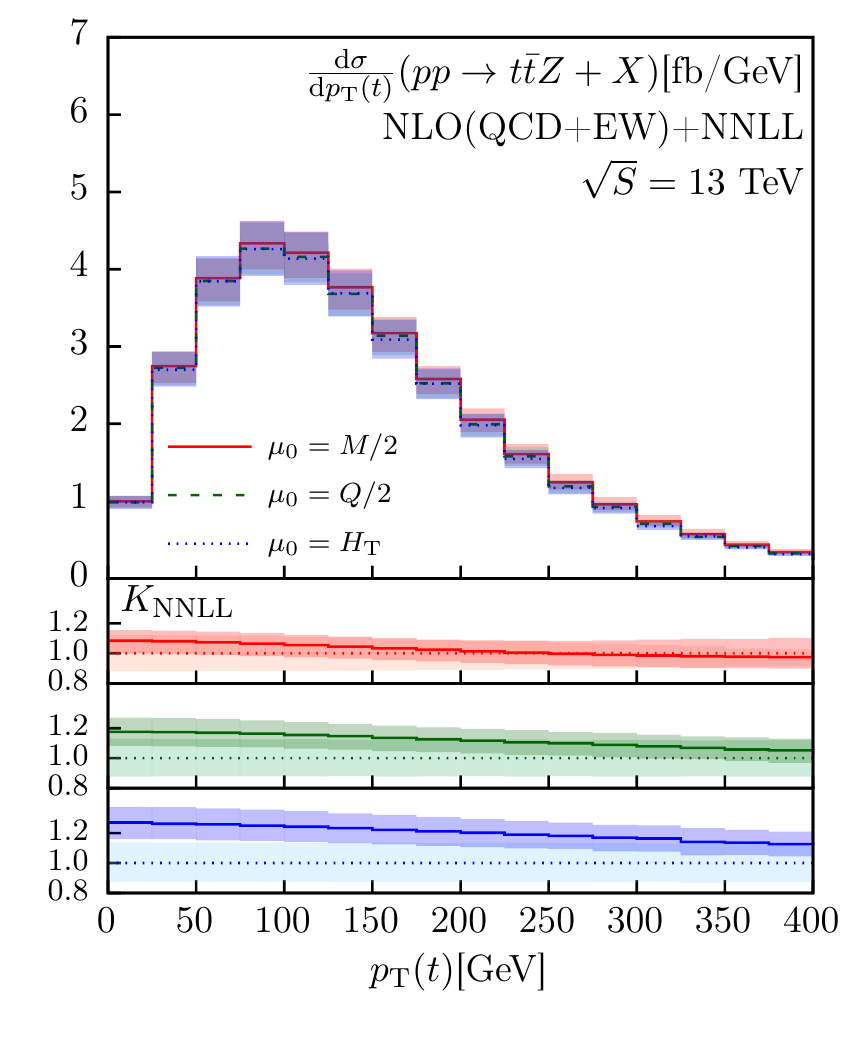}
\includegraphics[width=0.48\textwidth]{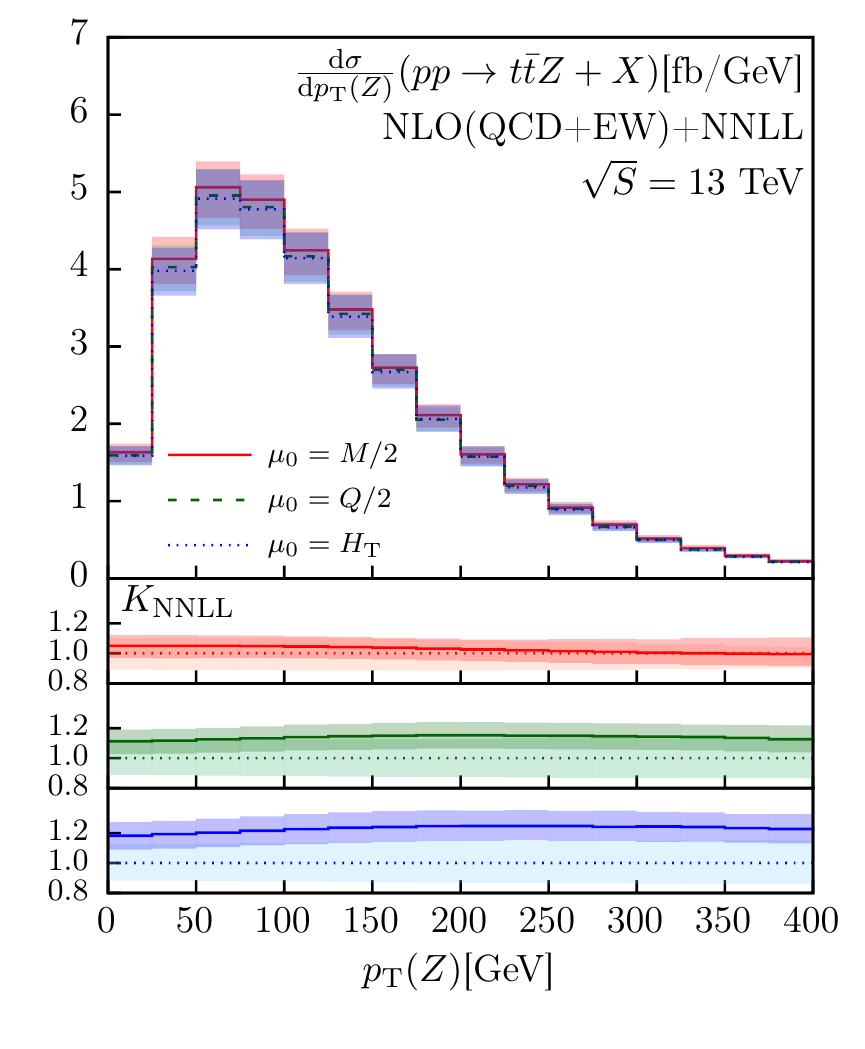}
\caption{The same as in Fig.~\ref{f:pTdiff_ttH} but for the $pp \to \ttZ$ process.} 
\label{f:pTdiff_ttZ}
\end{figure}

\begin{figure}[h!]
\centering
\includegraphics[width=0.48\textwidth]{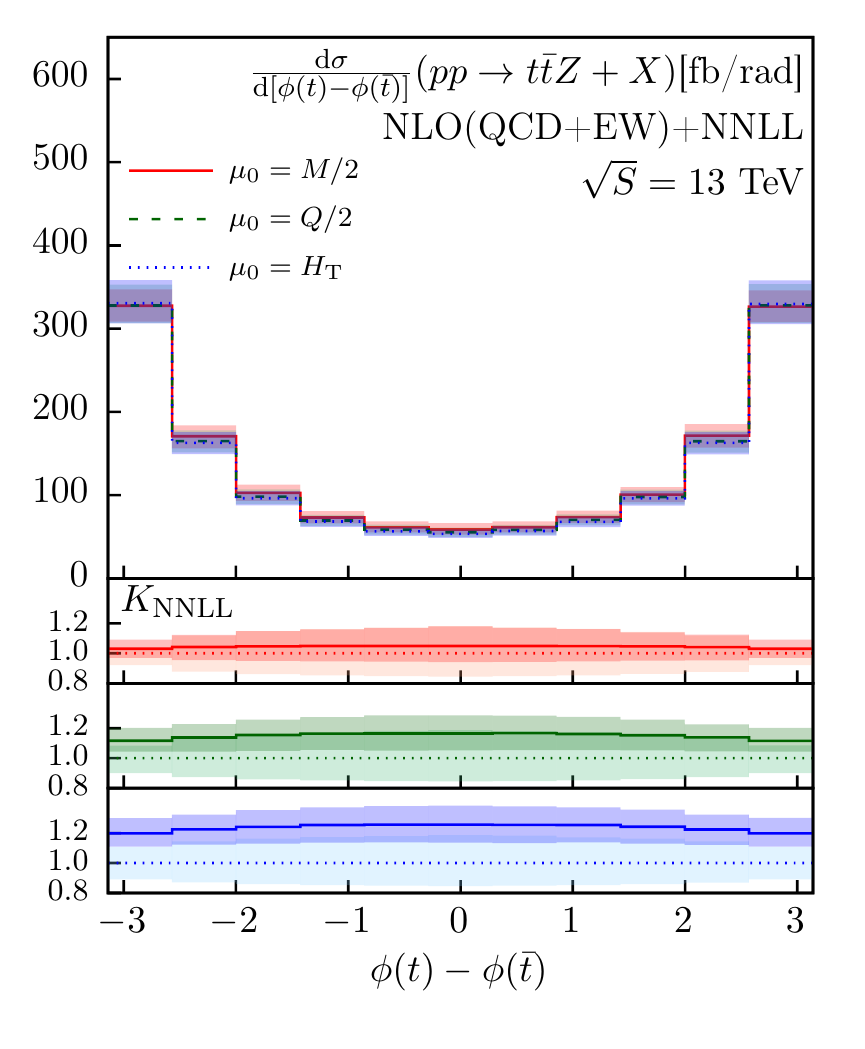}
\includegraphics[width=0.48\textwidth]{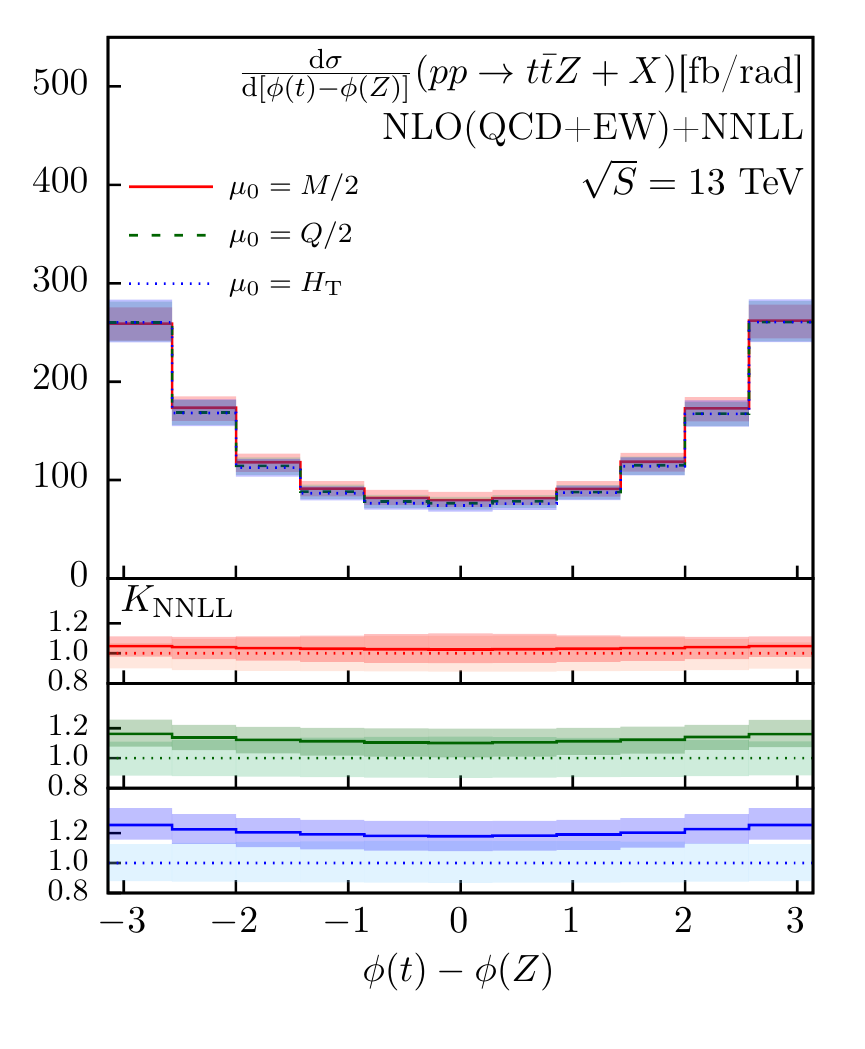}
\caption{The same as in Fig.~\ref{f:phidiff_ttH} but for the $pp \to \ttZ$ process.} 
\label{f:phidiff_ttZ}
\end{figure}

\begin{figure}[h!]
\centering
\includegraphics[width=0.48\textwidth]{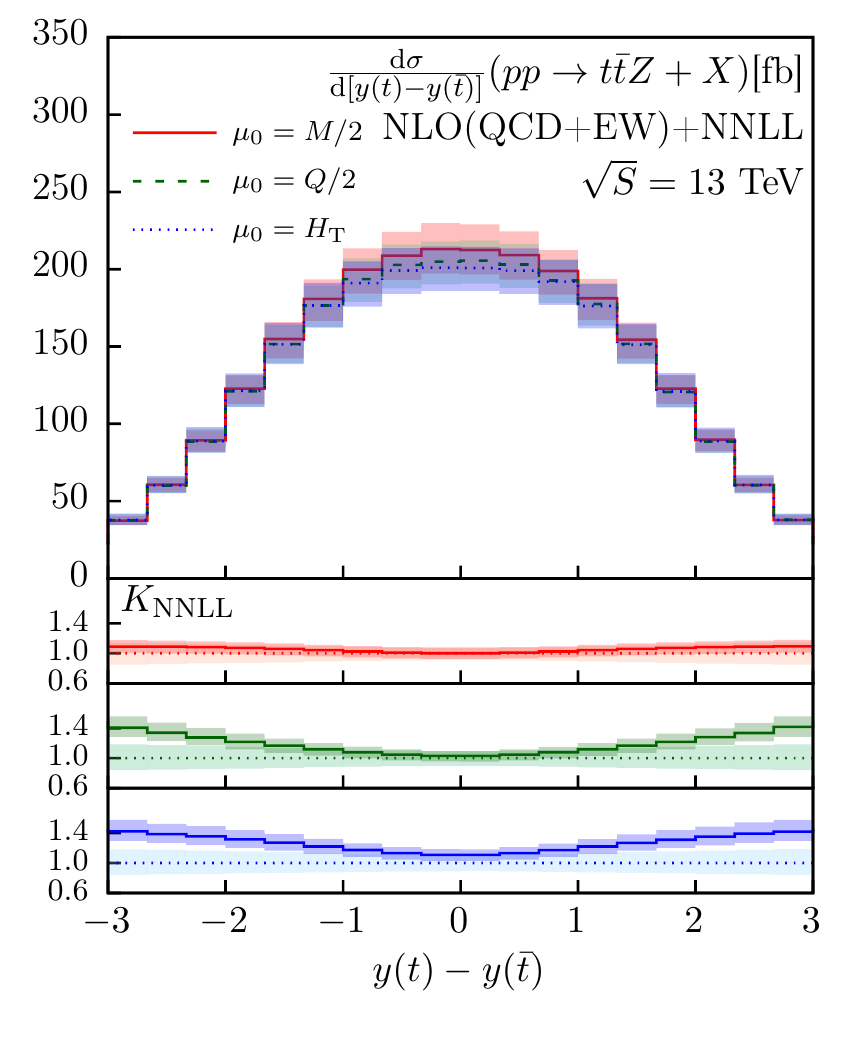}
\includegraphics[width=0.48\textwidth]{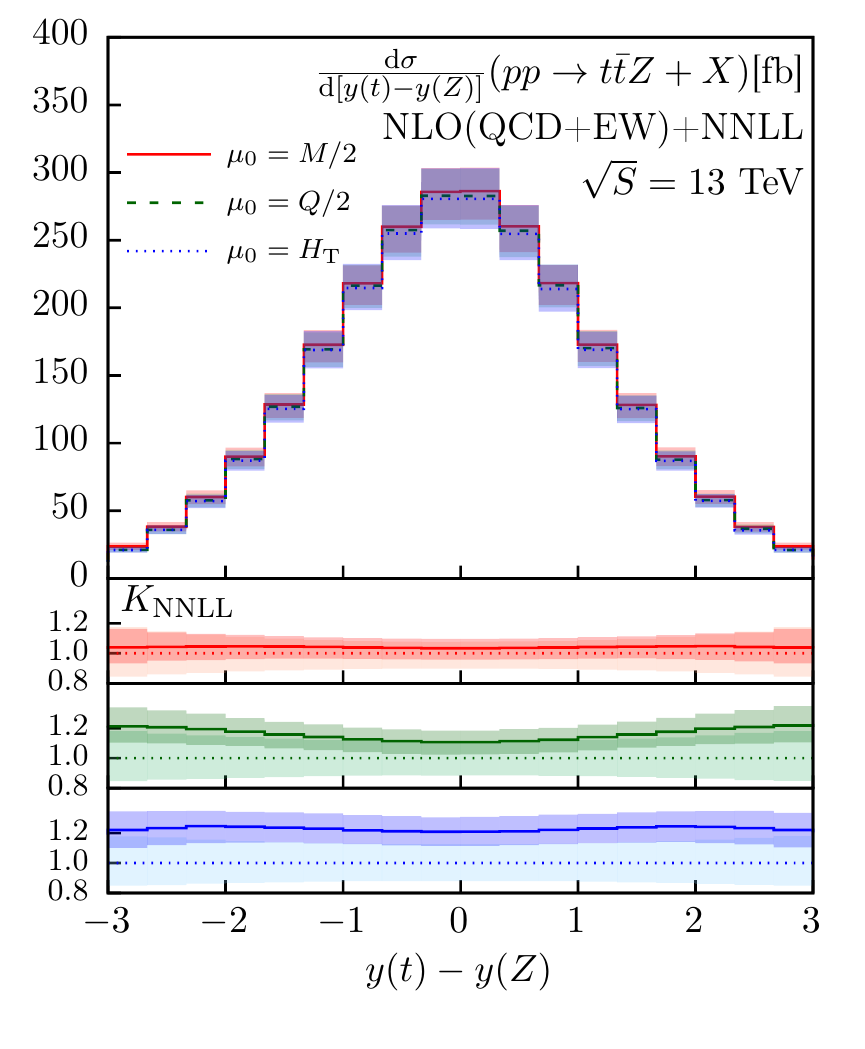}
\caption{The same as in Fig.~\ref{f:ydiff_ttH} but for the $pp \to \ttZ$ process.} 
\label{f:ydiff_ttZ}
\end{figure}

\begin{figure}[h!]
\centering
\includegraphics[width=0.48\textwidth]{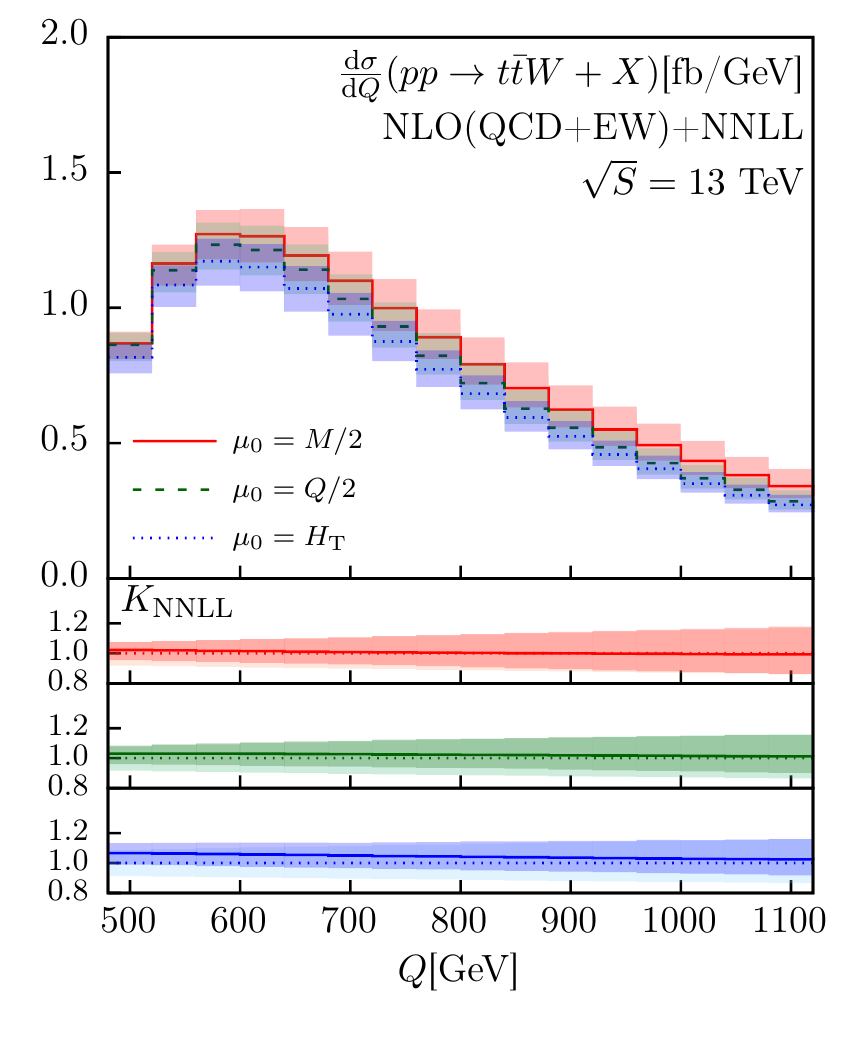}
\includegraphics[width=0.48\textwidth]{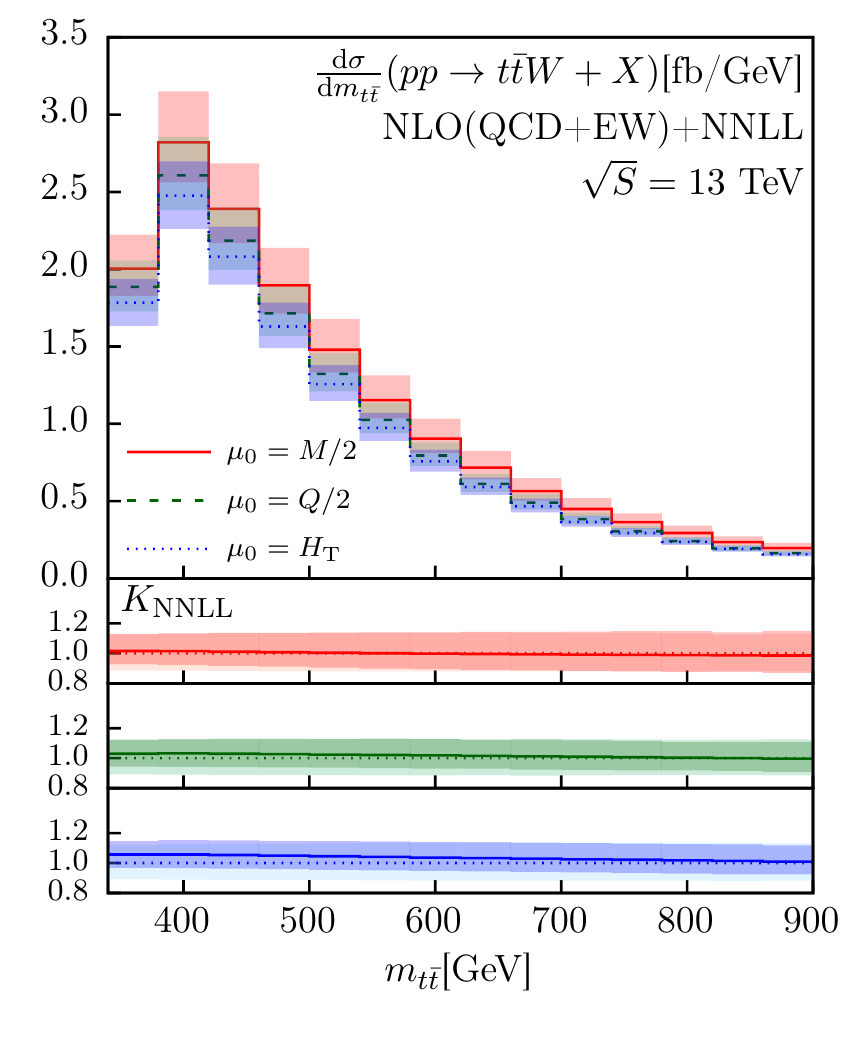}
\caption{The same as in Fig.~\ref{f:Qmttbardiff_ttH} but for the $pp \to \ttW$ process.} 
\label{f:Qmttbardiff_ttW}
\end{figure}

\begin{figure}[h!]
\centering
\includegraphics[width=0.48\textwidth]{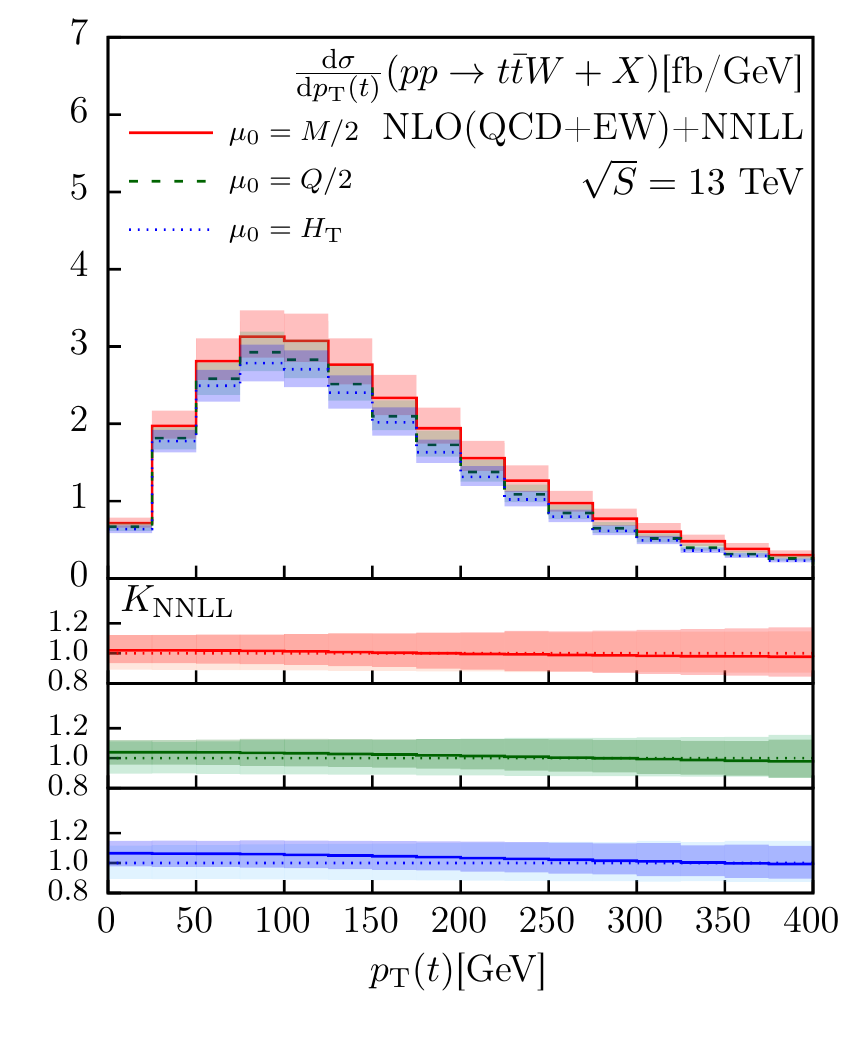}
\includegraphics[width=0.48\textwidth]{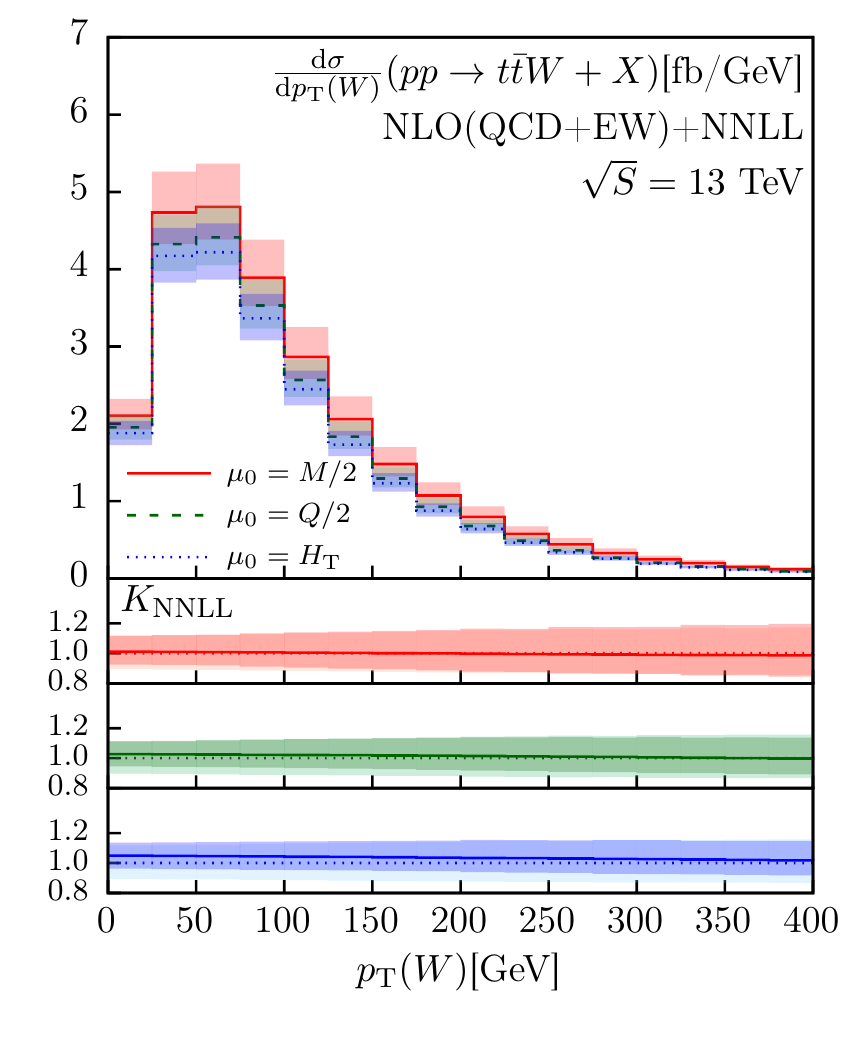}
\caption{The same as in Fig.~\ref{f:pTdiff_ttH} but for the $pp \to \ttW$ process.}
\label{f:pTdiff_ttW}
\end{figure}

\begin{figure}[h!]
\centering
\includegraphics[width=0.48\textwidth]{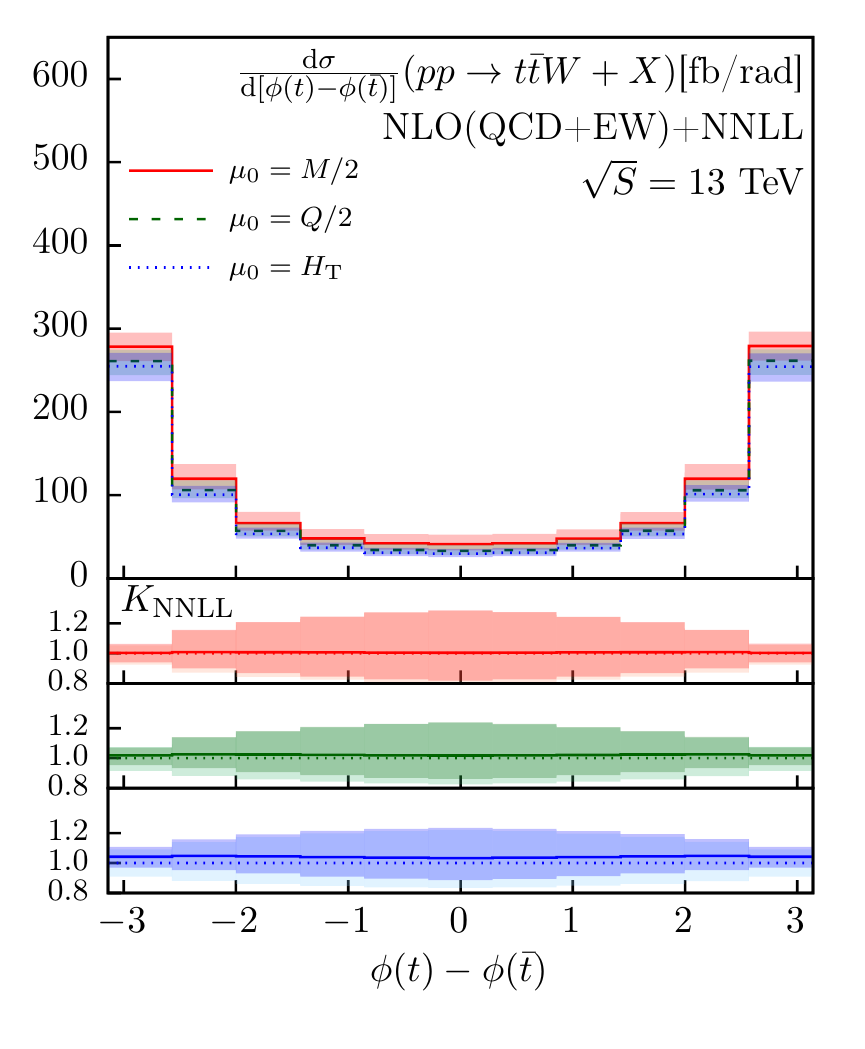}
\includegraphics[width=0.48\textwidth]{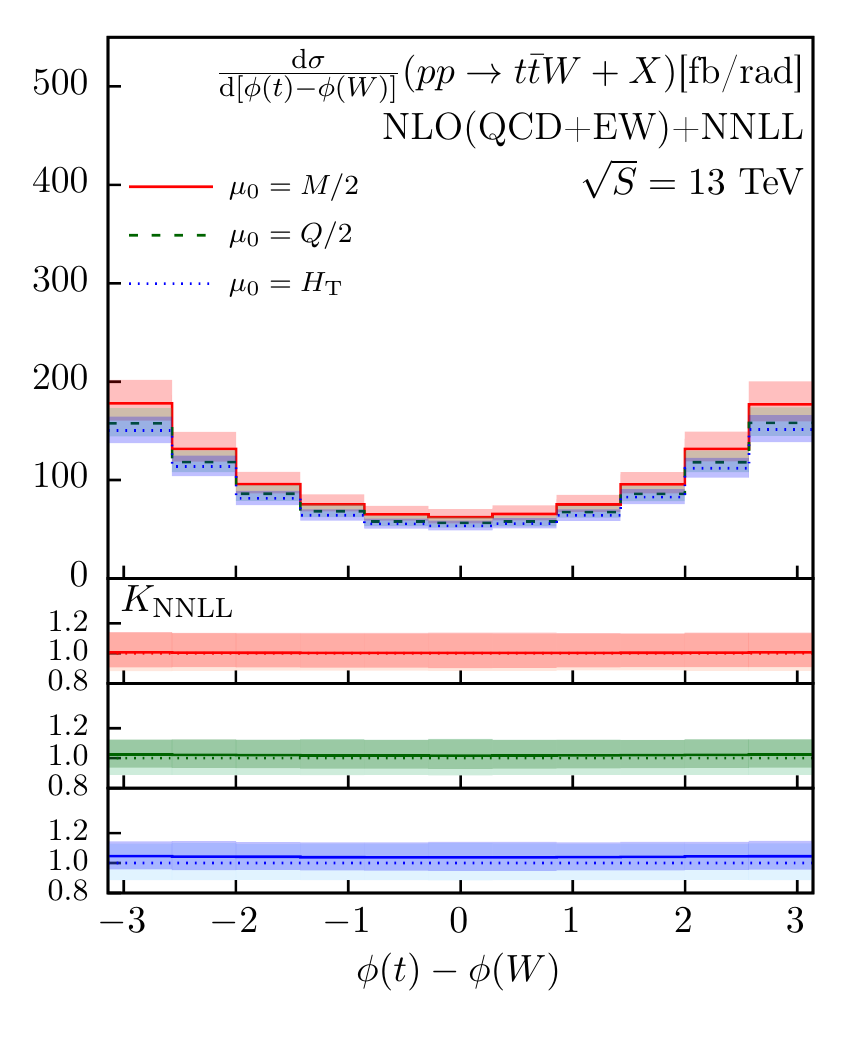}
\caption{The same as in Fig.~\ref{f:phidiff_ttH} but for the $pp \to \ttW$ process.} 
\label{f:phidiff_ttW}
\end{figure}

\begin{figure}[h!]
\centering
\includegraphics[width=0.48\textwidth]{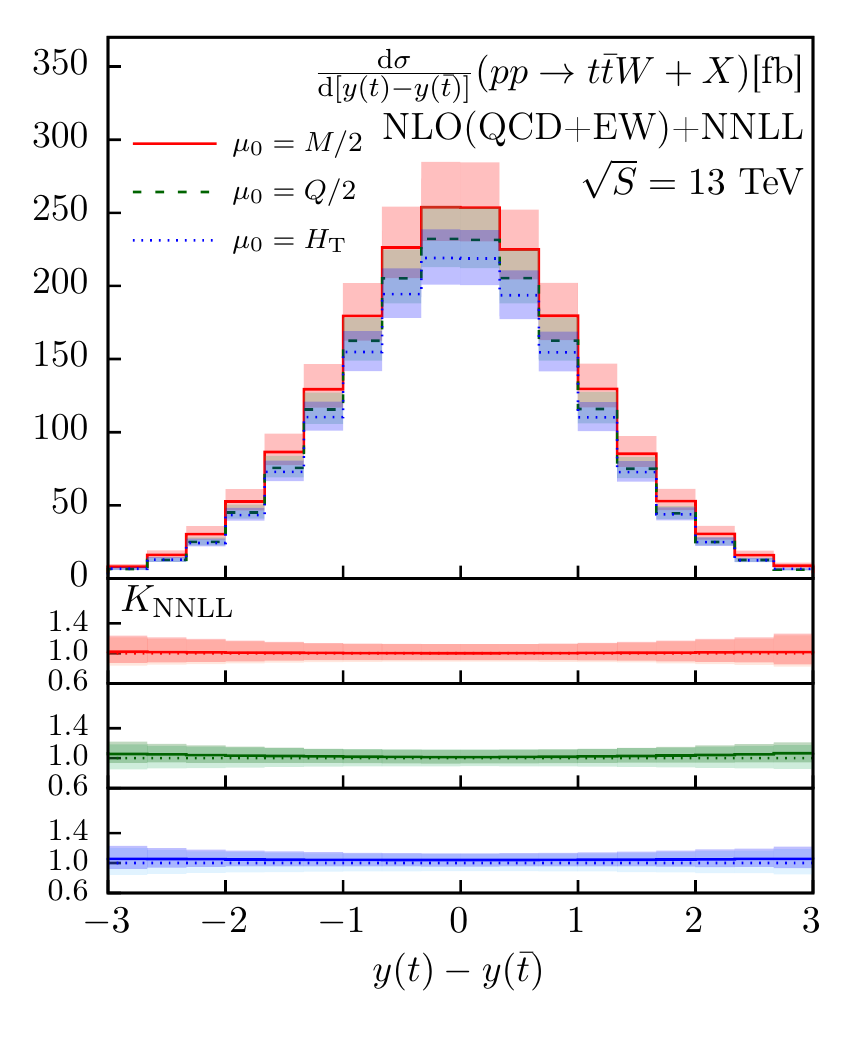}
\includegraphics[width=0.48\textwidth]{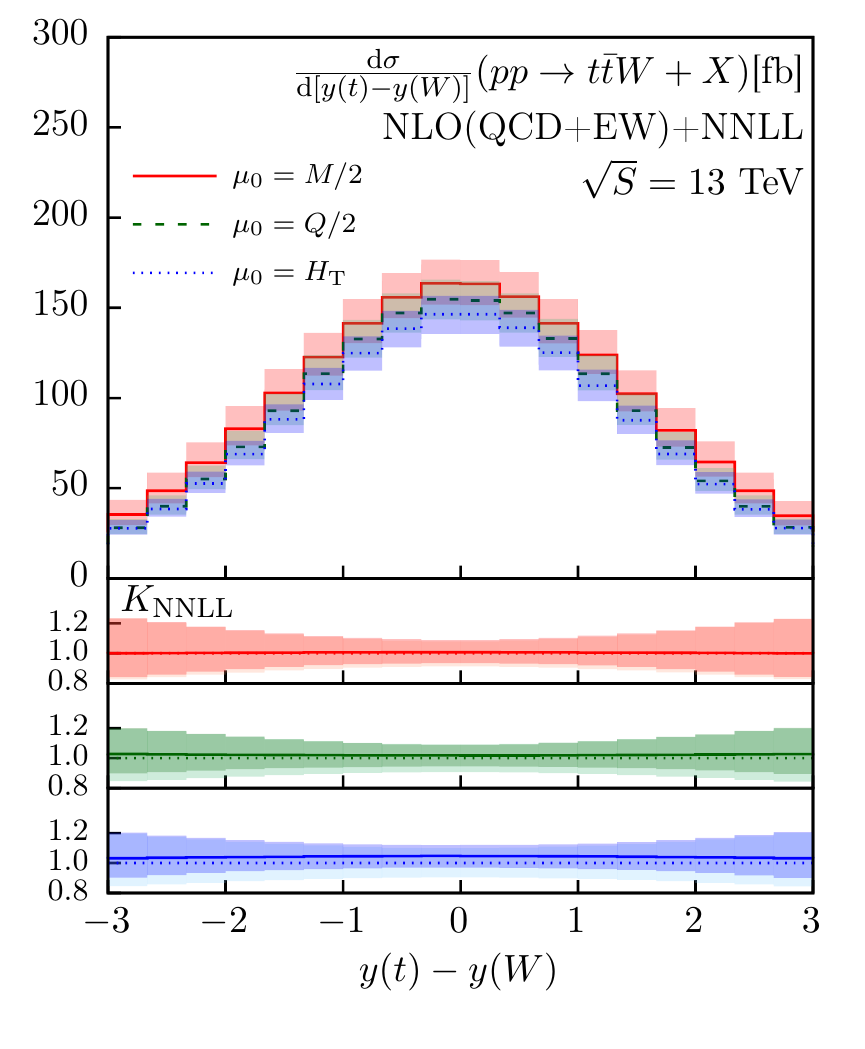}
\caption{The same as in Fig.~\ref{f:ydiff_ttH} but for the $pp \to \ttW$ process.} 
\label{f:ydiff_ttW}
\end{figure}

The NLO(QCD+EW)+NNLL distributions are presented in Figs.~\ref{f:Qmttbardiff_ttH}--\ref{f:ydiff_ttH} for the $\tth$ production, Figs.~\ref{f:Qmttbardiff_ttZ}--\ref{f:ydiff_ttZ} for the $\ttZ$ production and Figs.~\ref{f:Qmttbardiff_ttW}--\ref{f:ydiff_ttW} for the $\ttW$ production. We choose to show results for three representative scale choices: $\mu_0=M/2$, $\mu_0=Q/2$ and $\mu_0=H_T$. With the total cross section results for $\mu_0=H_T$ and $\mu_0=Q$ being very close, we expect  that such a choice of central scales covers the span of theoretical uncertainty for the predictions well. Results for NNLL differential distribution in $Q$, matched to NLO (QCD), were previously discussed by us in~\cite{Kulesza:2017ukk, Kulesza:2019lye} for the $\tth$ production and in~\cite{Kulesza:2018tqz} for the $\ttZ$ and $\ttW$ production. Here the NNLL results are matched to the NLO(QCD+EW) predictions and although there is not much difference in the behaviour of the $d \sigma / d Q$ cross section due to the presence of the EW corrections, we include the   $d \sigma / d Q$ cross section at NLO(QCD+EW)+NNLL  in our presentation for completeness.  The top panels of Figs.~\ref{f:Qmttbardiff_ttH} -- \ref{f:ydiff_ttZ} show an excellent agreement for the $\tth$ and $\ttZ$ NLO(QCD+EW)+NNLL predictions obtained for the three scale choices. As the impact of higher order logarithmic corrections is weaker for the $\ttW$ cross sections, the spread of predictions for $\ttW$ distributions does not get substantially decreased by adding NNLL resummation.  The relatively small effect of the NNLL corrections on the $\ttW$  distributions is in line with the behaviour of the total cross sections, see Fig~\ref{fig:totalxsec_13TeV_EW} and Ref.~\cite{Kulesza:2018tqz}. In contrast to the  $\tth$ and $\ttZ$ processes, the $ttW$ production at LO involves only the $q \bar q'$ channel. Correspondingly the NNLL resummation at leading power takes into account only the soft gluon emission from incoming quark lines which by means of  colour factors is much weaker than the emission from gluon lines.  The NNLL contributions are then simply too modest to outweigh the scale dependence of the NLO(QCD+EW) result to the same extent as they do for the $gg$ channel dominated processes, leading to a bigger spread in the central values of NLO(QCD+EW) NNLL predictions. This also explains why the reduction of the scale uncertainties for various central scale choices due to resummation is weaker for the  $ttW$ production, compared to the $\tth$ and $\ttZ$ processes.

The lower three panels in the figures show ratios of the NLO(QCD+EW)+NNLL distributions to the NLO(QCD+EW) distributions, i.e.\ the $K_{\text{NNLL}}$ factor, calculated for different values of $\mu_0$. The dark shaded areas indicate the scale errors of the  NLO(QCD+EW)+NNLL predictions, while light-shaded areas correspond to  the scale errors of the  NLO(QCD+EW) results. We observe that the ratios can differ substantially depending on the final state, observable or the central scale. Generally, the NNLL resummation has the biggest impact on the predictions obtained for $\mu_0=H_T$ among the three scale choice we study. In the case of the distributions in $Q$, $m_{t \bar t}$, $p_T(t)$, $p_T(B)$, $\phi(t) - \phi(\bar t)$, $\phi(t) - \phi(B)$ the ratios show that resummation can contribute as much as ca.\ 20\% (30\%) correction to the $\tth$ ($\ttZ$) distribution at this scale choice. As observed in ~\cite{Kulesza:2017ukk} and in ~\cite{Kulesza:2018tqz}, the size of the NNLL corrections to the invariant distribution in $Q$  mildly increases with $Q$. The same can be seen for the distribution in $m_{t \bar t}$, c.f.\ Figs.~\ref{f:Qmttbardiff_ttH} and~\ref{f:Qmttbardiff_ttZ}. The $p_T(t)$  distributions, on the other hand, receive the biggest NNLL corrections towards smaller values of $p_T$, whereas the corrections to the the $p_T(H)$, $p_T(Z)$ distributions get most pronounced for moderate values of $p_T$, see Figs.~\ref{f:pTdiff_ttH} and~\ref{f:pTdiff_ttZ}. From Figs.~\ref{f:phidiff_ttH} and~\ref{f:phidiff_ttZ} it can be observed that the distributions in the difference between the azimuthal angles of the top and the antitop quark are impacted the most for collinear configurations for almost all scale choices. The difference between azimuthal angles of the top and the $H$ or $Z$ boson show an opposite behaviour, i.e.\ they are enhanced for the back-to-back configurations in the transverse plane. The distributions in differences between rapidities, in particular  $y(t) - y(\bar t)$, can receive corrections of up to ca.\ 40\%, especially at high values of rapidity differences. These distributions also receive smaller NNLL corrections at lower rapidity differences, with the corrections generally growing as the difference in rapidity grow. For the reasons described above, the $\ttW$ distributions, on the other hand, get modified by a few percent, at most reaching up to~10\%, corrections.

The NNLL effects in the differential distributions are similar for the present paper and
Ref.~\cite{Broggio:2019ewu}. The comparison may be performed for the $Q$, $m_{t\bar t}$,  $pT(V)$ and $p_T(t)$ distributions. Within the ranges of the variables considered here, the overall picture is similar in both frameworks. The NNLL corrections do not affect the shapes of the distributions strongly and exhibit only mild kinematical dependence. The typical effects of NNLL are positive and of order 10\%. The NNLL effects lead to a better theoretical precision of the prediction. The improvement is
stronger for $\tth$ and $\ttZ$, resulting in the error band at or below 10\% level, and weaker in $\ttW$, where the uncertainty bands are close or above 15\%. 

Regarding the comparison with other results in the literature, in particular Ref.~\cite{Broggio:2019ewu}, the same remarks as in the discussion of the total cross section apply: different scale set-ups and choices do not allow to draw conclusions on the impact of formally subleading terms but we can examine absolute values of differential cross sections.
As an example, we compare the invariant mass $Q$ distributions obtained
here with the corresponding ones from Ref.~\cite{Broggio:2019ewu}. For
this purpose we apply the envelope method to the bin situated at the
peak of $Q$-distribution, at $Q=Q_{\text{max}}$. The
$K_{\text{NNLL}}(Q_{\text{max}})$ factors obtained this way are compared
with the values read off the plots in \cite{Broggio:2019ewu}. For the
$\tth$ production we obtain
$K_{\text{NNLL}}(Q_{\text{max}})=1.07^{+0.08} _{-0.08}$ which should be
compared with $1.02^{+0.08} _{-0.06}$  from \cite{Broggio:2019ewu}. For
the $\ttZ$ process we get $K_{\text{NNLL}}(Q_{\text{max}}) =1.13^{+0.09}
_{-0.10}$ compared to $1.05^{+0.11} _{-0.09}$ from
\cite{Broggio:2019ewu}. For the $\ttW$ process we get
$K_{\text{NNLL}}(Q_{\text{max}})=1.04^{+0.13} _{-0.12}$, whereas the
authors of \cite{Broggio:2019ewu} obtain $0.99^{+0.09} _{-0.09}$ for
$\ttW^+$.
\footnote{Since the $K$-factors for $\ttW^+$ and $\ttW^-$
in Ref.~\cite{Broggio:2019ewu} are very similar, we compare them with
our $K_{\text{NNLL}}$ for $\ttW$.} We see a similar agreement to the one found in comparison for the inclusive cross section. The size of the resummation corrections lie within the scale uncertainty of one another. In addition the size of the scale uncertainties in the different approaches is comparable, with the exception of $\ttW$.

\begin{figure}[h!]
\centering
\includegraphics[width=0.48\textwidth]{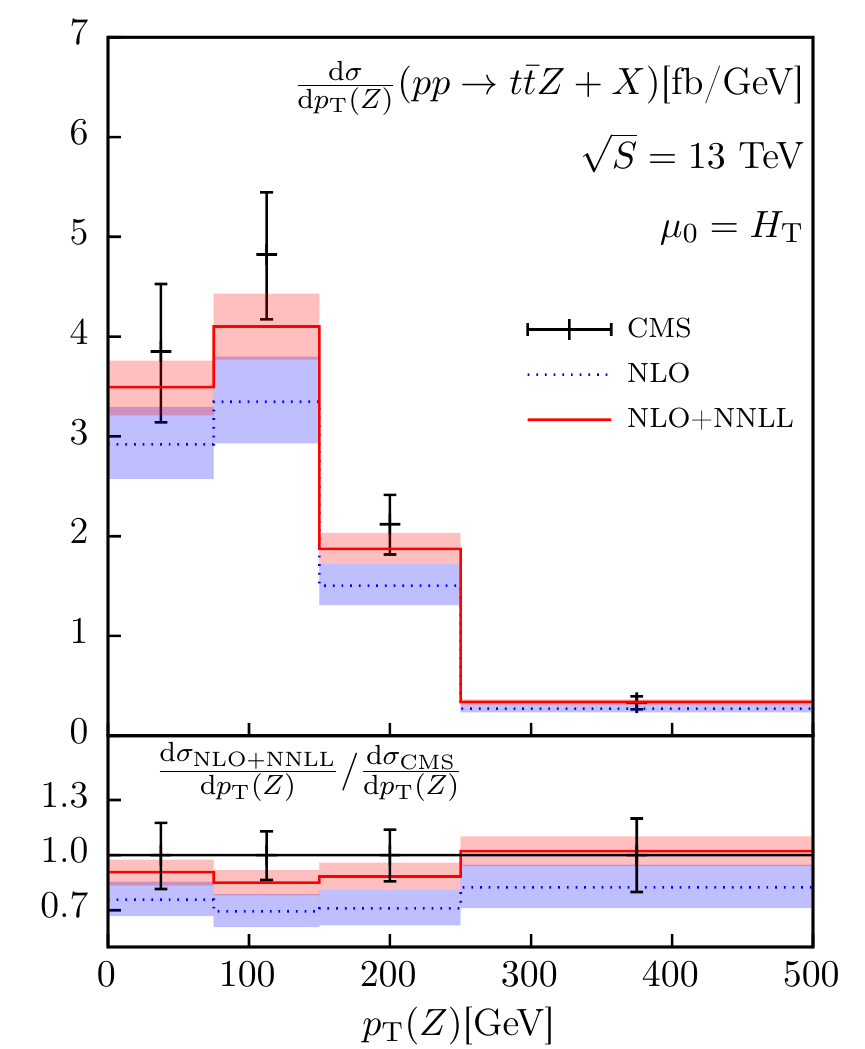}
\includegraphics[width=0.48\textwidth]{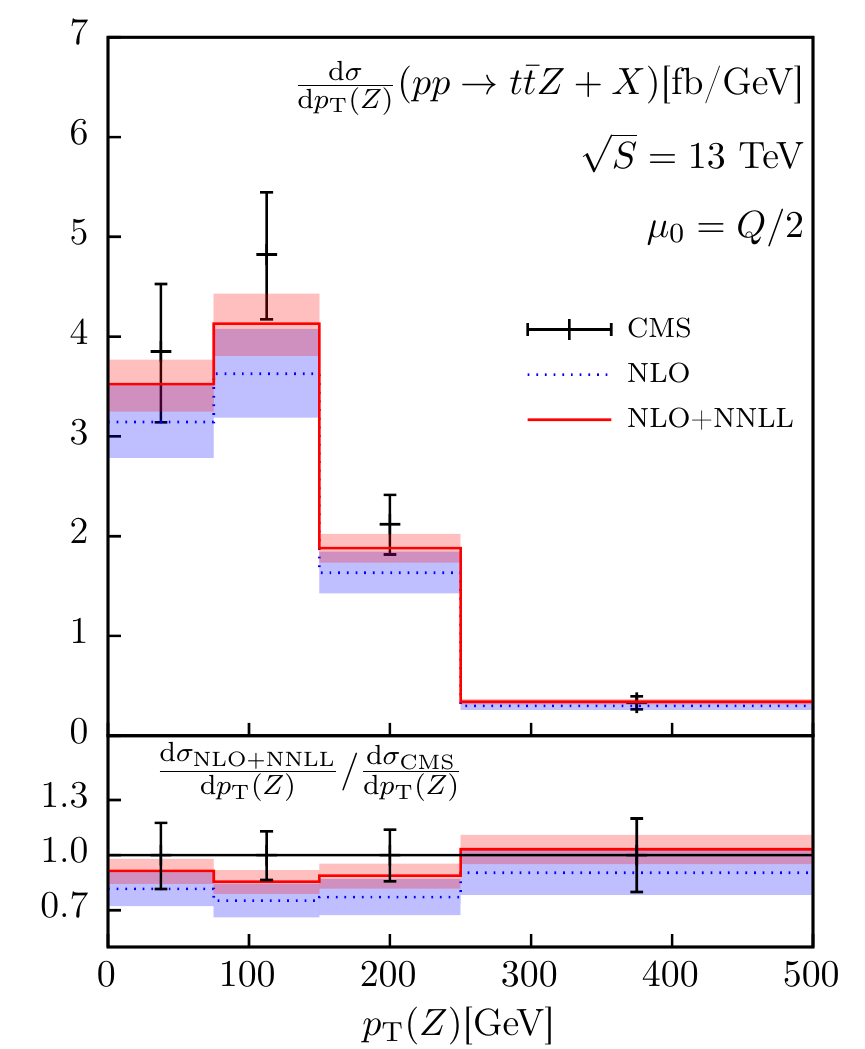}
\caption{Comparison of the  $p_T(Z)$  distribution measured by the CMS collaboration~\cite{CMS:2019too} with the NLO (QCD+EW)  and NLO (QCD+EW)+NNLL predictions for the central scale choices $\mu_0=H_T$ and $\mu_0=Q/2$. The shown theoretical uncertainty is from scale variation only. } 
\label{f:pTdiff_ttZ_CMS}
\end{figure}

\begin{figure}[h!]
\centering
\includegraphics[width=0.48\textwidth]{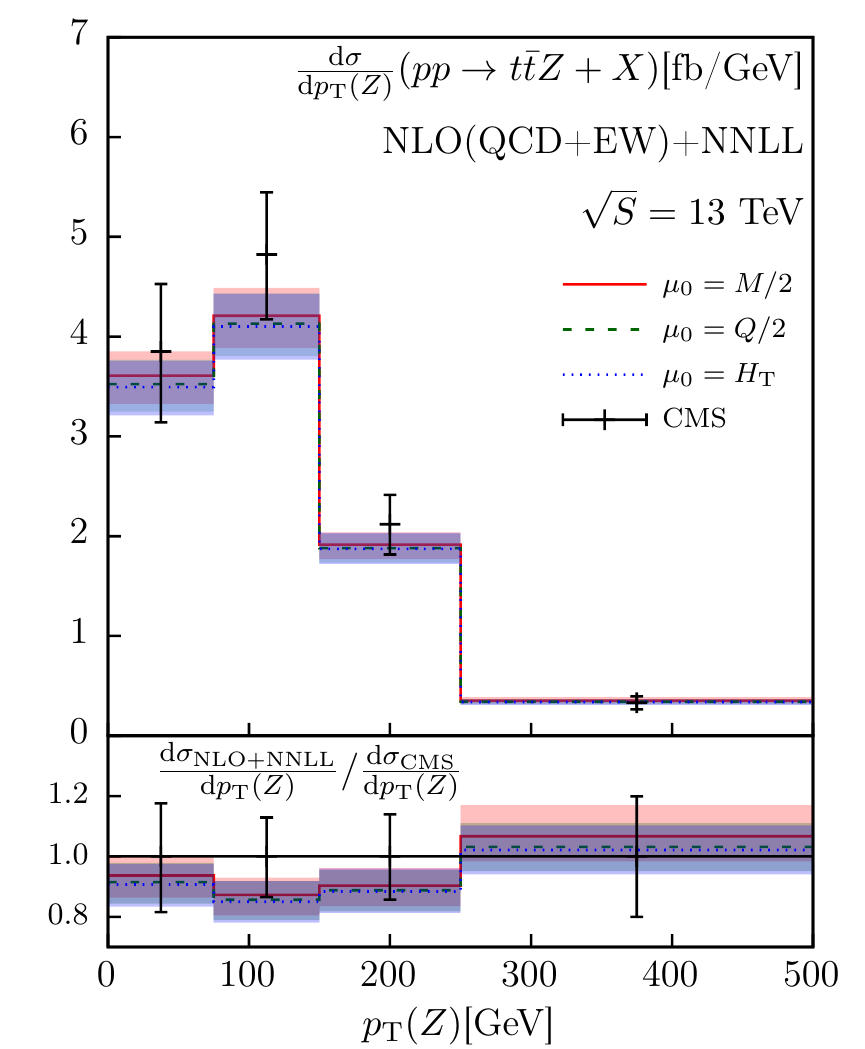}
\includegraphics[width=0.48\textwidth]{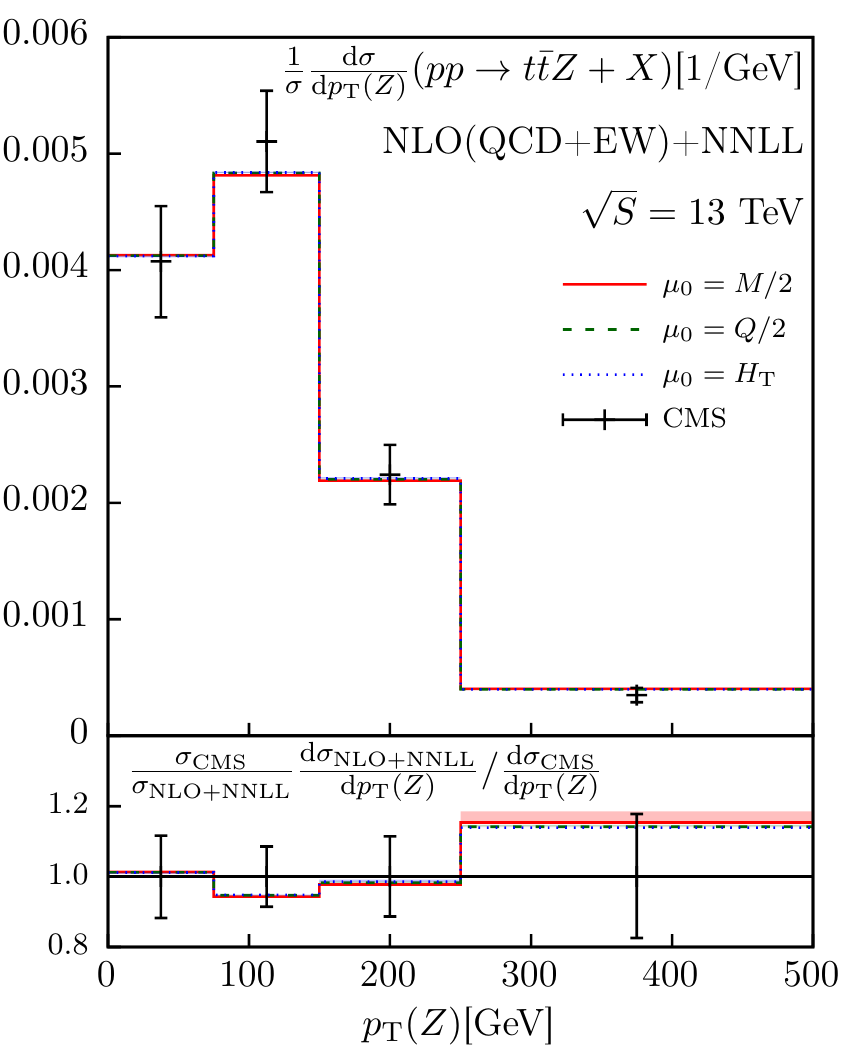}
\caption{Comparison of the $p_T(Z)$  distribution measured by the CMS collaboration~\cite{CMS:2019too} with the NLO (QCD+EW)+NNLL predictions for central scale choices considered in this paper (left) and the same comparison for the normalized distribution (right). The shown theoretical uncertainty is from scale variation only.} 
\label{f:pTdiff_ttZ_nnll_CMS}
\end{figure}

We finish the discussion by comparing our results for the differential distribution in $p_T(Z)$ with the very recent measurement of this distribution by the CMS collaboration~\cite{CMS:2019too}. In Fig.~\ref{f:pTdiff_ttZ_CMS} the NLO(QCD+EW) predictions are compared with the NLO(QCD+EW)+NNLL results for two different scales choices, $\mu_0=H_T$ and $\mu_0=Q/2$. We see that the resummed NNLL corrections bring the theoretical predictions closer to data and lead to a significant reduction in the scale dependence error.  The left plot in Fig.~\ref{f:pTdiff_ttZ_nnll_CMS} shows the comparison of the NLO(QCD+EW)+NNLL predictions for various scale choices, adjusted for the bin widths as used in the experimental measurement, while the right plot shows the same comparison for the shapes of the distributions.  In accordance with observations made above, the NNLL calculations yield our predictions remarkable stable w.r.t.\ the scale variation.

\FloatBarrier

\section{Summary}

In this paper we consider processes of associated top-antitop quark pair and a heavy boson $B = H,Z$ or $W^{\pm}$ production in $pp$ collisions. Theoretical predictions for total and differential cross sections at the LHC are obtained using the soft gluon resummation technique in Mellin space through the NNLL accuracy in QCD matched to existing NLO results in the QCD and the electroweak theory. The calculations are based on the framework developed in our earlier work
\cite{Kulesza:2015vda,Kulesza:2016vnq,Kulesza:2017ukk,Kulesza:2017jqv,Kulesza:2017hoc,Kulesza:2018tqz,Kulesza:2019lye, Kulesza:2019adl}, and the main aim of the present study is to provide an accurate theoretical reference for a wide set of observables that are or may be measured in the $pp \to \tth$, $pp \to \ttZ$ and $pp \to t\tb W^{\pm}$ scattering at the LHC. The framework applied offers the currently best available theoretical precision, with the reduction of the theoretical uncertainties due to scale variation reaching up to a factor of about two with respect to the corresponding NLO(QCD+EW) estimates. 

The main focus of the present study are the differential cross sections. Hence we present the 
distributions of the $t \tb B$ and $t \tb$ invariant masses, $p_T$ of the boson and $p_T$ of the top quark. Moreover we obtain the distributions for the azimuthal angle $\phi$ and rapidity $y$ differences: $\phi(t) - \phi(\tb)$, $\phi(t) - \phi(B)$, $y(t) - y(\tb)$ and $y(t) - y(B)$ for all the
considered bosons. The soft gluon resummation effects are found to be significant in the differential cross sections: they affect both the overall normalisation and the shapes. For dynamical scale choices, the magnitude of the NNLL corrections is up to 20--30\% of the NLO(QCD+EW) results, but in some kinematic regions the relative
NNLL contribution reaches 40\%. In general, the estimated theoretical uncertainty of the NLO(QCD+EW)+NNLL results is reduced w.r.t.\ the NLO(QCD+EW) predictions in all distributions. In particular, the resummation
greatly reduces dependence on the central scale choice. The strongest improvement of the theoretical accuracy is found for  $pp \to \tth$, $pp \to \ttZ$ processes, where two gluon fusion partonic channel is important. The improvement is only moderate for $pp \to t \tb W^{\pm}$, where the two gluon channel does not contribute below the NNLO accuracy.   

The theoretical estimates are compared to results of the recent CMS measurement \cite{CMS:2019too} of $d\sigma / dp_T(Z)$ in $pp \to \ttZ$ at $\sqrt{S} = 13$~TeV. The inclusion of soft gluon resummation is shown to significantly improve the agreement between the theoretical predictions and the experimental results, and the theoretical uncertainty due to the central scale choice is nearly completely eliminated in the NLO+NNLL results.

\section*{Acknowledgments}
We are grateful to R. Schoefbeck and J. Knolle for providing us data and very useful exchanges regarding the CMS analysis. This work has been supported by the DFG grant KU3103/2. AK would like to acknowledge the hospitality of and the support from the Erwin Schr\"odinger International Institute for Mathematics and Physics of the University of Vienna, where part of this work was carried out. The support of the Polish National Science Center (NCN) grants No.\ 2017/27/B/ST2/02755 and 2019/32/C/ST2/00202 are gratefully acknowledged.
VT acknowledges funding from the European Union's Horizon 2020 research
and innovation programme as part of the Marie Sk\l odowska-Curie Innovative
Training Network MCnetITN3 (grant agreement no. 722104).

\providecommand{\href}[2]{#2}\begingroup\raggedright\endgroup

\end{document}